\documentclass[11pt]{article}
\usepackage{epsfig}

\newcommand{\be}{\begin{eqnarray}}
\newcommand{\ee}{\end{eqnarray}}

\def\fr{\frac{1}{2}}

\def\mref#1{(\ref{#1})}

\def\bd{\begin{displaymath}}
\def\ed{\end{displaymath}}

\def\ba#1{\begin{array}{#1}}
\def\ea{\end{array}}
\def\nn{\nonumber}
\newfont{\Bbb}{msbm10 scaled 1200}

\begin{document}

\pagestyle{empty}

\begin{center}

{\LARGE\bf Borel Summable Solutions to 1D
Schr\"odinger Equation\\[0.5cm]}

\vskip 12pt

{\large {\bf Stefan Giller{$\dag$} and Piotr Milczarski{$\ddag$}}}

\vskip 3pt

Theoretical Physics Department II, University of {\L}\'od\'z,\\
Pomorska 149/153, 90-236 {\L}\'od\'z, Poland\\ e-mail: $\dag$
sgiller@krysia.uni.lodz.pl \\ $\ddag$
jezykmil@krysia.uni.lodz.pl
\end{center}
\vspace{6pt} 
\begin{abstract}
It is shown that so called fundamental solutions the
semiclassical expansions of which have been established earlier
to be Borel summable to the solutions themselves appear also to
be the unique solutions to the 1D Schr\"odinger equation having
this property. Namely, it is shown in this paper that for the
polynomial potentials the Borel function defined by the
fundamental solutions can be considered as the canonical one.
The latter means that any Borel summable solution can be
obtained by the Borel transformation of this unique canonical
Borel function multiplied by some $\hbar$-dependent and Borel summable
constant. This justifies the exceptional role the fundamental
solutions play in 1D quantum mechanics and completes the relevant 
semiclassical theory relied on the Borel resummation technique and 
developed in our other papers.

\end{abstract}

\vskip 3pt
\begin{tabular}{l}
{\small PACS number(s): 03.65.-W , 03.65.Sq , 02.30.Lt , 02.30.Mv} \\[1mm]
\end{tabular}

\newpage

\pagestyle{plain}

\setcounter{page}{1}

\section*{I.  Introduction}

\hskip+2em The semiclassical approximation is one of the most 
widely used approximate methods in physics, particularly, in 
quantum mechanics. In fact it is not limited only to pure power
 series expansions in the Planck constant $\hbar$ but it is used 
also in all problems which can be formulated semiclassically. 
The method can be applied in this way to, say, the quartic 
oscillator perturbation theory from the one hand \cite{1,4} and 
to a variety of problems with so called large-N expansions from the 
other \cite{27,28,2}. Therefore, independently of the expansion 
parameter we shall consider all such asymptotic series expansions 
as semiclassical. 

The method can be stated in the Schr\"odinger wave function 
formulation of quantum mechanics \cite{21,26} as well as in 
the Feynman path inegral form of the latter \cite{29,30}. 
Its main ingredient as the approximation method is to represent 
considered quantities by a limited number of first terms of the 
corresponding infinite series expansions, knowing usually that 
the series is typically asymptotic i.e. divergent. Therefore, 
contrary to the case of convergent series, such a representation 
of the expanded quantities is of a rather limited value. 

First, it can not be done arbitrarily accurate by enlarging a number 
of kept terms i.e. such an approximation can be only the best one in 
which the case the finite rest is exponentially small (in the parameter
 of the expansion) in comparison with the main contribution. 

Secondly, 
it is just these exponentially small differences which can become 
dominating in other domains of the expansion parameter being arbitrarily
 close to the original one i.e. such a finite representation of quantities
 by their corresponding asymptotic series are strongly limited to the 
original domain and it can not give any information about the analytic
 properties of the quantities considered as functions of the expansion
 parameter. In particular this finite sum can not be continued 
analytically outside the original domain of obtaining it. 

The latter means that the considered quantities can not be
 recovered in some simple way by the knowledge of a finite 
number of terms of their semiclassical expansions even if 
the series are abbreviated at their least terms. (In the 
latter case the approximation is considered to be best). 
In fact these analytic properties are determined rather by
 the behaviour of so called large order terms of the series.
 It is just the properties of these large order terms (considered
 as functions of their order) which allow us actually to reconstruct 
quantities represented by such semiclassical series. In particular 
if these terms grow with their order not faster than factorially  
then the Borel method of summation of the diverging series can be 
used in such cases. We shall call such semiclassical series Borel summable.

In the case of the Borel summable semiclassical expansions the Borel 
method of summation can be used in the following way. 

First the Borel function is determined approximately after the knowledge 
of a limited number of first terms of the asymptotic series expansion of 
the considered quantity. Namely, by its definition the approximate Borel 
function is obtained as a sum of these known first terms divided by the 
corresponding factorials. The sum represents in this way the abbreviated 
Taylor series expansion corresponding to the exact Borel function. (The 
latter is obtained if all the terms of the asymptotic series are used to
 construct the Taylor series in the above way). The last series has a 
finite radius of convergence and therefore an additional knowledge even
 approximate of the singularity structure of the corresponding Borel 
function is still necessary for an approximate recovery of this function 
from its abbreviated Taylor series. These knowledge can be extracted to
 the known extent from the detailed knowledge of the large order behaviour
 of the considered semiclassical series. Having (or assuming) however this 
knowledge a function with the desired singularity structure can be 
constructed and its Taylor series expansion can be compared with the known 
abbreviated Taylor series so that the free parameters of the assumed 
singularities can be determined. Finally by the Borel transformation of
 the Borel function the original quantity is reproduced approximately in 
this way \cite{28}. 

The above reproduction can still be performed in the spirit of 
asymptotic expansions and on different levels of accuracy. The 
lowest level is obtained when the Borel integral is substituted 
by its best asymptotics. It means that the quantity considered is 
represented again by a finite sum a definite number of first terms 
of which coincide exactly with the original terms used to construct
 the (approximate) Borel function. However this sum can now contain
 much more terms since it ends on the least term of the asymptotic 
series corresponding to the Borel integral. Therefore it can approximate 
the considered quantity  much better than the sum of the original terms 
the method has started with.

However, since the method contains still additional information 
about the singularity structure of the Borel function then one step 
further can be done in getting still better level of accuracy by 
extracting from the Borel integral so called exponentially small 
contributions. Such computations are known as the hyperasymptotic 
ones \cite{11,31}. The finite semiclassical sum is then completed by
 the exponentially small contributions the forms and numbers of which
 are determined by the known (assumed) singularity structure of the 
Borel function \cite{6}. The latter means that the Borel summability 
allows us to realize the {\it principle of resurgence} i.e. to recover the 
information contained in the divergent tails of the semiclassical 
series \cite{11,6,12,13,14,15,20}. 

 It should be noted also, however, that the exponentially small 
contributions are of their own importance since in many cases of 
quantities considered these contributions are dominant. Among the 
latter cases the most well known one is the difference between the 
energy levels of different parities in the symmetric double well \cite{21}. 
But these are also the cases of transition probabilities in the tunnelling
 phenomena \cite{21} or their adiabatic limits in the time dependent 
problem of transitions between two (or more) energy levels 
(see \cite{16,17} and references cited there) or the exponential 
decaying of resonances in the week electric field (see \cite{18,19}
 and references cited there).

The applicability of the Borel resummation to the semiclassical 
expansions in quantum mechanics has been proved by many authors 
\cite{32,33,7}. Particularly, the 1-D quantum mechanics offers a
 possibility of constructing a full semiclassical theory relied 
on the Borel resummation \cite{1,6,10}. Namely, several years ago
 one of the authors of the present paper discovered \cite{1} that
 for a large family of analytical potentials including all the 
polynomial ones there are solutions to 1D stationary Schr\"odinger
 equation for which their well defined semiclassical expansions 
are Borel summable to the solutions themselves. These solutions 
appearing for polynomial potentials in a finite number were called 
fundamental because of their completeness for solving any one-dimensional 
problem \cite{2}. Their Borel summability property played an essential role 
in many of their applications \cite{1}. In particular this property allowed
 us to prove the Borel summability of energy levels for most of the 
polynomial potentials.

On the other hand it is easy to construct solutions to the Schr\"odinger
 equation (in fact, infinitely many of them) with well defined Borel 
summable semiclassical expansions but with results of such Borel 
resummations not coinciding with the initial solutions generating
 the series. However the results of the Borel resummations are again 
solutions to the Schr\"odinger equation since in general each successful 
Borel resummation of any semiclassical series always leads to some 
solution to the Schr\"odinger equation.

In this paper we want to demonstrate an exceptional role the fundamental
 solutions mentioned above play with respect to the Borel summability 
property showing that they provide a general scheme for a construction 
of Borel summable solutions to the 1D stationary Schr\"odinger equation at
 least for polynomial potentials. A main ingredient of such a scheme is 
an observation that the Borel function of some fundamental solution is 
not only such a function for any other fundamental solution but it is 
also a Borel function allowing us to construct any Borel summable solution
 to a given 1D Schr\"odinger equation with polynomial potential. 

The latter conclusion means that in all the semiclassical problems 
in 1-D quantum mechanics in which the Borel resummation method is 
to be applied the fundamental solutions should be used preferably. 

Our way of
considering the problem of the Borel summability in 1D quantum
mechanics makes use of the global features of the fundamental
solutions and the Stokes graphs related to them and as such is
to some extent complementary to the way utilized by Delabaere $et \; al$ 
\cite{12,13} making use of rather local properties of the
considered quantities.  

Our method can be also used to analyze
the adiabatic limits considered by Joye $et \; al$ \cite{16} at least in
the case of two energy levels \cite{adia}. The cases of several levels need,
however, a generalization of our method since these cases are
described by systems of the linear equations in numbers larger than two.  

To make the original results of our paper more
transparent we have formulated them in many places in the forms
of theorems or lemmas equipped with the corresponding proofs.
However we do not consider our paper to pretend to a full formal
mathematical rigor supposing most of the presented ideas to be
sufficiently obvious and clear by presented proofs or when
confronted with our earlier papers or with the papers of other
authors mentioned.  

The paper is organized as follows. 

In the next section we remind a construction and basic properties of
the fundamental solutions and Borel functions corresponding to
them as well.  

In Sec. 3 we show that the Borel functions
corresponding to different fundamental solutions are only
different branches of the same unique Borel function and can be
recovered by the Borel transformations performed along suitably
chosen paths on the 'Borel plane'. We show also here that there
are two ways of integrations in the Borel plane providing us
with the Borel summable solutions to the Schr\"odinger equation
which, on their own, coincide each, up to $\hbar$-dependent
multiplicative constants, with the corresponding fundamental solutions.  

In Sec. 4 we consider in some details a general
expression for the semiclassical expansions to the Schr\"odinger
equation and introduce there also their standardized forms. We
point out in this section an essential difference between the
forms of the latter for the Borel summable and non-summable quantities.  

In Sec. 5 we show the existence almost at each
point of the $x$-plane two pairs of the base solutions to the
Schr\"odinger equation with well defined Borel summable
semiclassical asymptotic but not summed to the solutions
themselves. The semiclassical expansions of the solutions
considered in this section and their Borel resummations are a
particular illustration of our main thesis that a result of any
such a resummation is always some fundamental solution. 

 In Sec. 6 we generalize the results of Sec. 4 and show that the Borel
function defined by the fundamental solutions can be considered
as canonical in a sense that up to a multiplicative $\hbar$-dependent
constant any Borel summable solution to the Schr\"odinger equation
can be obtained by the Borel transformation of this canonical
Borel function. This means that each Borel summable solution has
to be essentially some of the fundamental solutions.  

Sec. 7 is a discussion of the results of the paper.

\section*{II.  Fundamental 
solutions to 1D stationary Schr\"odinger \\ equation}

\vspace{12pt}

\hskip+2em Let us remind shortly basic lines in defining fundamental 
solutions \cite{1,2}.

A set of fundamental solutions is attached in a unique way to a so called 
Stokes graph corresponding to a given polynomial potential $V(x)$ of 
n{\it th} degree. Each Stokes graph is a collection of lines (Stokes lines) in
the complex $x$-plane which are a loci of points where the real
parts of action functions defined by the following $n$ integrals:
\be
W_i(x,E) = \int_{x_{i}}^{x} \sqrt{q(y,E)}dy \nn
\label{1}
\\
\\q(x,E) = 2V(x) - 2E\nn 
\ee
vanish. In (\ref{1}) $E$ is the energy of the system and $x_i$,
$i=1,2,...,n$, are roots of $q(x)$.  

The fundamental solutions are
defined in infinite connected domains called sectors with
boundaries of the latter consisting of Stokes lines and $x_i$'s, see Fig.1a. 

In a sector $S_k$ a corresponding fundamental solution $\psi_k$ to the
Schr\"odinger equation: 
\begin{eqnarray}
\psi^{\prime\prime}(x) - \hbar^{-2} q(x) \psi(x) = 0
\label{2}
\end{eqnarray}
has Dirac's form: 
\begin{eqnarray}
\psi_{k}(x) = q^{-\frac{1}{4}}(x){\cdot}
e^{\frac{\sigma_k}{\hbar} W_k(x)}{\cdot}{\chi_{k}(x)} 
\label{3}
\end{eqnarray}

\vskip 15pt
\begin{tabular}{cc}
\psfig{figure=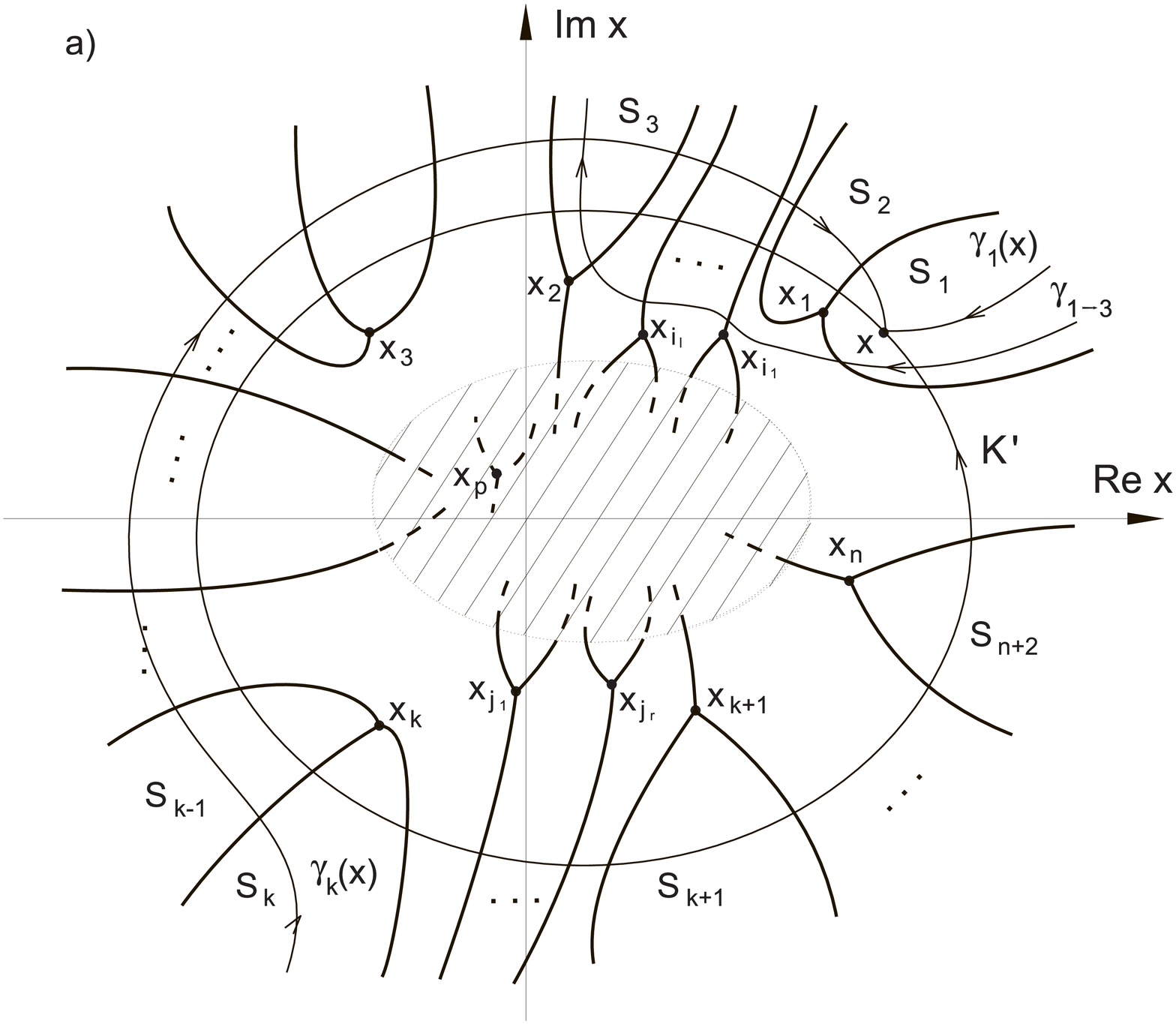,width=8cm}&\psfig{figure=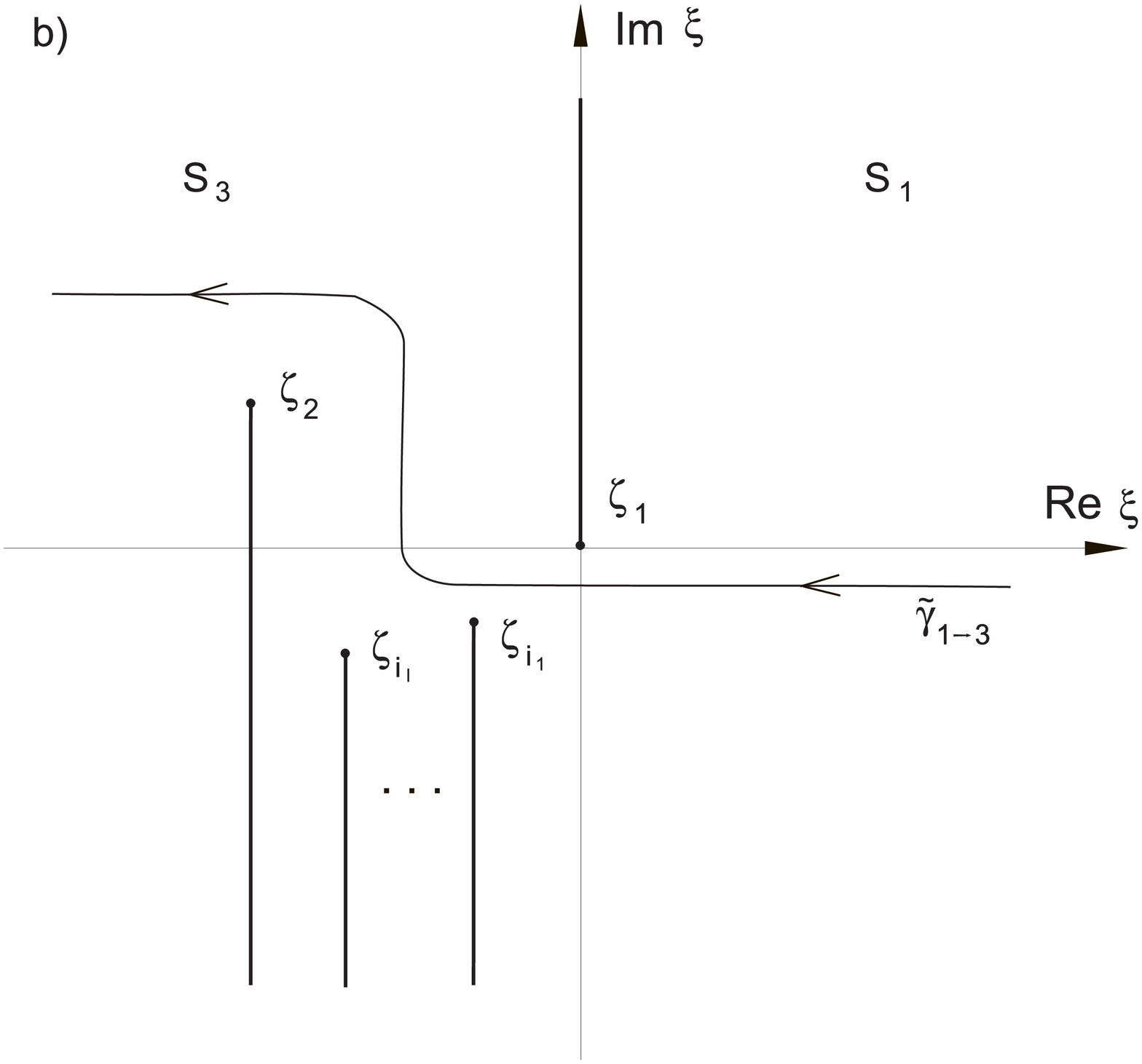,width=7cm} 
\\
Fig.1a A general form of the Stokes & Fig.1b The $\xi$-variable
map of Fig.1a
\\
graph for the polynomial potential & containing the sectors $S_1$ and $S_3$
\\
\\
\end{tabular}
with $x_k$ lying at
the boundary of $S_k$ and with a sign $\sigma_k (= \pm 1)$, (which we shall 
call a $signature$ of the solution (\ref{3})) chosen in such a way to have:
\begin{eqnarray}
 \Re \left(\sigma_k{W_k}(x) \right) < 0
\label{4}
\end{eqnarray}

The amplitude factor $\chi_{k}(x,\hbar)$ in (\ref{3}) has the following 
Fr\"oman and Fr\"oman's form \cite{3}:

\begin{eqnarray}
\chi_{k}(x,\hbar) = 1 + \sum_{n{\geq}1}
\left( \frac{{\sigma_k} \hbar}{2} \right)^{n} \int_{\infty_k}^{x}dy_1
\int_{\infty_k}^{y_1}dy_2 \ldots 
\int_{\infty_k}^{y_{n-1}}dy_n
\omega(y_1)\omega(y_2) \ldots \omega(y_n) 
{\times}
\label{5}
\end{eqnarray}
\begin{eqnarray*}
\left( 1 -
e^{-\frac{2\sigma_k}{\hbar}(W_{k}(x)-W_{k}(y_1))} \right)
\left(1 - e^{-\frac{2{\sigma_k}}{\hbar}(W_{k}(y_1)-
W_{k}(y_2))} \right)
\cdots \left(1 - e^{-\frac{2{\sigma_k}}{\hbar}{(W_{k}(y_{n-1})-
W_{k}(y_n))}} 
\right) 
\end{eqnarray*}
with
\begin{eqnarray}
\omega(x) = {\frac{1}{4}}{\frac{q^{\prime\prime}(x)}{q^{\frac{3}{2}}(x)}} - 
{\frac{5}{16}}{\frac{q^{\prime 2}(x)}{q^{\frac{5}{2}}(x)}} = - 
{q^{- \frac{1}{4}}(x)} \left( {q^{- \frac{1}{4}}(x)} \right)^{\prime{\prime}}
\label{6}
\end{eqnarray}
and with integration paths in (\ref{5}) chosen to be canonical
\cite{1,2} i.e. on such paths the following condition is satisfied:
\begin{eqnarray}
\sigma_k \Re \left(W_{k}(y_j) - W_{k}(y_{j+1}) \right) \geq 0
\label{7}
\end{eqnarray}
for any ordered pair of integration variables (with $y_0 = x$). The
condition (\ref{7}) ensures the solution (\ref{3}) to vanish at the infinity
$\infty_k$ of the sector $S_k$.  

A domain $D_k (\supset S_k)$ where $\chi_k (x)$ can be
represented by (\ref{5}) with the canonical integration paths is
called canonical. In each $D_k$ the following semiclassical
expansion for $\chi_k (x)$ can be deduced from (\ref{5}) by standard methods
(see also the next section):
\begin{eqnarray*}
\chi_k (x,\hbar) \sim \chi_{k}^{as} (x,\hbar) =
\sum_{n\geq{0}} 
\left(\frac{\sigma_k {\hbar}}{2} \right)^{n} \chi_{k,n}(x) 
\end{eqnarray*}
\begin{eqnarray}
\chi_{k,n}(x) =\int_{\infty_k}^x dyq^{-\frac{1}{4}}(y)\left(q^{-\frac{1}{4}
(y)}\chi_{k,n-1}(y)\right)^{\prime\prime}= \int_{\infty_{k}}^{x}dy_{n}q^{-\frac{1}{4}}(y_{n}) 
\label{8}
\end{eqnarray}
\begin{eqnarray}
\times\left( q^{-\frac{1}{4}}(y_{n}) 
\int_{\infty_{k}}^{y_{n}}dy_{n-1}q^{-\frac{1}{4}}(y_{n-1}) 
\left( {\ldots} q^{-\frac{1}{4}}(y_{2}) 
\int_{\infty_{k}}^{y_2}dy_{1}q^{-\frac{1}{4}}(y_{1})
\left( q^{-\frac{1}{4}}(y_{1}) \right)^{\prime{\prime}} \ldots 
\right)^{\prime{\prime}} \right)^{\prime{\prime}}\nn
\\ \nn
\\n = 1,2,\ldots,\;\;\;\;\;\;\;\;\;\;\;\;\; \chi_{k,0}(x)\equiv 1\;\;\;\;\;\;\;\;\;\;\;\;\;\;\;\;\;\;\;\;\;\;\;\;\;\;\;\;\;\;\;\;\;\;\;\;\;\nn 
\end{eqnarray}

What has been said above assumed real and positive value of
$\lambda \equiv \hbar^{-1}$ (we prefer to use rather $\lambda$ as a more 
convenient variable). However when considering Borel summability properties 
of $\chi_k(x,\lambda)$ it is unavoidable to complexify $\lambda$. 
If it is done the only change in the above descriptions of properties of 
Fr\"oman and Fr\"oman solutions to Schr\"odinger equation is to substitute 
$W(x,E)$ in the conditions (\ref{4}) 
and (\ref{7}) by $e^{\imath \phi} W(x,E)$ where $\phi = arg \lambda$ . 
Of course, the domains $D_k$ as well as Stokes graph
itself depend then on $\phi$. In particular, all the Stokes lines rotate then
around the corresponding turning points $x_i$, $i=1,2,...,n$, they emanate
from by the angle $-2\phi/3$. For $\phi = \pm \pi$ Stokes graph comes back to 
its  initial position i.e. a dependence of Stokes graph on $\phi$ is 
periodic with $\pi$ as its period. Such a full rotation of Stokes graph 
we shall call {\it cyclic}. We can
use the cyclic rotations to enumerate all the sectors according
to the order they come into each other by the subsequent cyclic
rotations starting from the one chosen arbitrarily. We shall
assume from now on such a convention for the sector ordering
with the numbers attached to sectors increasing anticlockwise.

By a cyclic rotation a solution $\psi_k (x,\lambda)$ from a sector $S_k$
transforms into a solution $\psi_{k-1}(x,\lambda)$ or $\psi_{k+1}(x,\lambda)$ 
(modulo $n+2$, the last number being the total number of sectors for a
polynomial potential of the n{\it th} degree) according to whether the
rotation of Stokes graph is clockwise or anticlockwise respectively. Of
course, for a fixed $x$ after at most two subsequent cyclic
rotation (in the same direction) the path of integration in (\ref{5})
stops to be canonical if it was as such before the rotation
operations. Let us note also that 
making, say clockwise, $n+2$ subsequent cyclic rotations a solution
$\psi_k (x,\lambda)$ does not come back exactly to its initial form
(\ref{3}) but acquires rather an additional phase factor 
which in the case of even $n$ is equal to
$(-\imath)^n \exp(\lambda \sigma_k \oint_{K'} \sqrt{q(x,E)} dx)$
where the (closed) contour $K'$ encloses (anticlockwise) all $n$ roots of
the potential $V(x)$ (see Fig.1a). 
In the case of odd $n$ one needs to surround all the roots
twice as much to close the corresponding path of analytical
continuation of $\chi_k(x,\lambda)$ in the $x$-plane with the result 
analogous with the even case. It means of course that as a fuction of
$\lambda$ a solution $\psi_k (x,\lambda)$ branches infinitely around the
points $\lambda =0,\; \infty$ of the $\lambda$-plane \cite{1}.  

As we have mentioned earlier it was shown in \cite{1} that in sector $S_k$
the series (\ref{8}) can be Borel summed to $\chi_k (x,\lambda)$ itself. 
To be a little bit detailed it was shown in \cite{1} that when $x \in S_k$ : 

$1^0$ the size of a sector in the $\lambda$-plane where the expansion 
(\ref{8}) is valid is larger than $2\pi$; and 

$2^0$ the rate of grow of $\chi_{k,n}(x)$ in (\ref{8}) with $n$ is factorial.

The last property which was established by
an application of the Bender-Wu formula \cite{4} ensured that the
following Borel series: 
\begin{eqnarray}
\sum_{n \geq 0} \chi_{k,n} (x) \frac{(-s)^n}{n!}
\label{9}
\end{eqnarray}
was convergent in a circle: $\mid s \mid < \mid W_k (x,E) \mid$.

On its turn the property $1^0$ above ensured that the series (\ref{9}) define
Borel functions ${\tilde{\chi}}_k (x,s)$ holomorphic in the halfplane: 
$\Re s < -\sigma_k \Re {W_k}(x,E)$
allowing to recover ${\chi}_k (x,\lambda)$ from the series (\ref{8}) by the 
following Borel transformation of ${\tilde{\chi}}_k (x,s)$: 
\begin{eqnarray}
\chi_k (x,\lambda) = 2\lambda \int_{C_{\phi}}e^{2\lambda s} 
{\tilde{\chi}}_k (x,s) ds
\label{10}
\end{eqnarray}
where $C_{\phi}$ is a halfline in the Borel halfplane 
$\Re s < -\sigma_k \Re {W_k}(x,E)$ starting at the infinity and ending at $s=0$ 
with $\phi$ as its declination angle $(\pi/2 \leq \phi \leq 3\pi/2)$.

However, for the latter transformations to exist it is neccessary for 
the functions $\tilde{\chi}_k (x,s)$ to be holomorphic only in some vicinity 
of a ray $\arg s = \phi_0$ along which the transformation (\ref{10}) can be 
taken \cite{5}. Such a limiting situation appears when $\chi_k (x,\lambda)$
is continued from the sector $S_k$ to other domains of Stokes graph so that
such a continuation generates singularities of ${\tilde{\chi}}_k (x,s)$
in the half plane $\Re s < 0$ close to the ray $\arg s = \phi_0$. A mechanism
of such singularity generations has been described by one of the present 
authors \cite{6}. Some of these singularities are fixed and the others are
moving with their positions in the $s$-plane depending on $x$. The 
possibility to perform the Borel transformation (\ref{10}) of 
${\tilde{\chi}}_k (x,s)$ along the ray $\arg s = \phi_0$ to get ${\chi}_k 
(x,\lambda)$ disappears at the moment when two of the moving singularities 
which are localized close to the ray $\arg s= \phi_0$ pinch the latter.
It is clear that such cases depend continuously on $x$ i.e. for a given 
$\phi_0$ in the
domain $D_k(\phi_0)$ of the $x$-plane there is a maximal domain 
$B_k(\phi_0)$ $(D_k(\phi_0) \supset B_k(\phi_0) \supset S_k(\phi_0) )$ 
inside which the series (\ref{8}) for ${\chi}_k (x,\lambda)$ is Borel 
summable 
along $C_{\phi_0}$ to ${\chi}_k (x,\lambda)$ itself. To find a boundary of 
$B_k(\phi_0)$ one can use Voros' technique \cite{7} of rotating of the 
reduced Stokes graph (i.e. the one obtained in the limit 
$\mid \lambda \mid \rightarrow \infty$) 
with changing of $\arg \lambda$ (see also \cite{1}): when 
$x \in {\partial}B_k(\phi_0)$ the total change of $\arg \lambda$ preserving 
the canonicness of the integration path in (\ref{5}) running from 
$\infty_k(\phi_0)$ to $x$ cannot be greater than $\pi$. Let us note also that 
for $x \in B_k(\phi_0)$ but close to  $x_0 \in {\partial}B_k(\phi_0)$
the Borel transformation of ${\tilde{\chi}}_k (x,s)$ along the ray 
$\arg s = \phi_0$ provides us with ${\chi}_k (x,\lambda)$ defined for 
$\pi/2 - \phi_0 \leq \arg \lambda \leq 3\pi/2 - \phi_0$.

\section*{III.  Properties 
of the Borel functions ${\tilde{\chi}}_k (x,s)$}

\hskip+2em In this section we shall establish properties of the
fundamental solutions and their corresponding Borel functions
not discussed in our papers quoted in the previous sections.

First let us note that we  can drop the subscribe $k$ at the Borel functions 
${\tilde{\chi}}_k (x,s)$ because in fact all these 
functions define one and the same Borel function.
This property is the subject of the following theorem.

\newtheorem{moje}{Theorem}
\begin{moje} 

Let $\tilde\chi(x,s)$ coincide with $\tilde\chi_1(x,s)$ when 
$x \in S_1$ and $|s| < |\xi(x)|$ and where $\xi (x)\equiv -\sigma_1W_1 (x,E)$. 
Then 

a) ${\tilde{\chi}} (x,s)$ coincides with the Borel functions
${\tilde{\chi}}_k (x,s),\; k=2,...,n+2$, corresponding to the remaining 
fundamental solutions;

b) each fundamental solution can be obtained from ${\tilde{\chi}} (x,s)$
when $x \in S_1$ by the Borel transformation with the 
integration path obtained by a suitable homotopic deformation of the path 
$C_1$ used to recover ${\chi}_1 (x,\lambda)$.  
\end{moje}
 
\vskip 6pt

$Proof$.  

The part (a) of the theorem follows directly from the definitions of 
the Borel functions $\tilde{\chi}_k(x,s)$ by (\ref{9}). Namely, for 
$x\in K'\cap S_1$ 
whilst $|s | < | \xi(x)|$, 
we can transform the 
coefficients ${\chi}_{1,n}(x)$ defining $\tilde{\chi}(x,s)$ by the series 
(\ref{9}) into the corresponding 
coefficients ${\chi}_{k,n}(x)$. To do it, it is enough to 
continue
analytically the infinite limit $\infty_1$ of all the integrations in 
(\ref{8}) from the sector $S_1$ to the sector $S_k$ to achieve the infinity 
$\infty_k$ of the sector $S_k$. 
Of course this is a deformation of the integration path $\gamma_1(x)$ 
in (\ref{8}) into the $\gamma_k(x)$ one
but this does not affect the integrations if $none$ of the turning points is
touched by the deformed path what is assumed and seen on Fig.1a. 
In other words such 
a deformation should be homotopic. Due to this operation
we have, of course, ${\chi}_{1,n}(x)\equiv {\chi}_{k,n}(x), n\geq 1$ and, by
(\ref{9}) $\tilde{\chi}_k(x,s)\equiv \tilde{\chi}(x,s)$, $k=2,...,n+2$.

Before going to the proof of the part (b) of the theorem let us make the 
following comments to the proof done so far.

First let us call as the standard paths the integration paths $\gamma_k(x)$ 
which appear 
in this way in (\ref{8}) linking the infinity 
$\infty_k$, $k=2,...,n+2$, with the point $x$, $x\in K'\cap S_1$ . 

Let us note further that continuing the infinite tail of the standard path 
by moving by the subsequent sectors around the $closed$ contour $K'$ 
(to close the contour $K'$ in the case of the odd $n$ it is necessary
to round the infinity point two times) we have to come back again to 
the sector $S_1$ when crossing the 'last' $S_{n+2}$ one. The consistency 
condition which follows then from (\ref{8}) demands that the integral 
$\oint_{K'}q^{-\frac{1}{4}}(x)(q^{-\frac{1}{4}}(x)\chi_{1,n}(x))''dx$ should 
then vanish for each $n \geq 0$. One can easily check that it happens 
indeed for all the polynomial potentials.

For the factor ${\chi}_{k}(x,\lambda)$, however, as given by (\ref{5})
if $k \neq 2,n+2$ the standard path is of course 
non canonical (see Fig.1a). But this means merely that ${\chi}_k (x,\lambda)$
obtained in this way is an effect of its analytical continuation along $K'$
from the sector $S_k$, 
where it could be initially defined, to the sector $S_1$. This of 
course means also that ${\chi}_{k}(x,\lambda)$ cannot be obtained from 
(\ref{10}) by the integration along a halfline but rather by the corresponding 
integration along some more complicated path described below.
  
To restore, however, the Borel function corresponding to 
${\chi}_{k}(x,\lambda)$ when $x\in S_k$ it is necessary only to continue 
${\tilde{\chi}} (x,s)$ analytically by moving the point $x$ from the sector 
$S_1$ to $S_k$ along the contour $K'$. Of course at the end of this 
continuation the standard path linking $\infty_k$ with the continued $x$
is then found completely in the sector $S_k$ being there a
typical canonical path for the integrations in (\ref{5}) and (\ref{8}). 

On the Borel plane the latter analytical continuation of $\tilde{\chi}(x,s)$ 
corresponds to a motion of the branch point of at $s= \xi (x)= 
-\sigma_1\int_{x_1}^{x} \sqrt{q(y,E)}dy$  shown in Fig.3
along the line $\tilde K'$ which is the image of $K'$ on the $s$-Riemann 
surface given by the map $s=\xi(x)$. During this motion the argument of this 
branch point changes by $(k-1)\pi$.

Let us finish these comments of the part (a) of the theorem by noticing that 
this part can be proved also using the results 
of the discussion performed in Appendix 1 and preceding Theorem 6.

The validity of the part (b) follows easily just from Theorem 6 of Appendix 1.
Namely, to prove this part consider first ${\tilde{\chi}}(x,s)$ when $x$ is 
continued from the sector $S_1$ to the sector $S_3$ along the contour $K'$ 
shown in Fig.1a whilst $s$ is kept fixed. A pattern of the first 
sheet of the $s$-Riemann surface is then shown on Fig.10b. To recover 
${\chi}_3(x,\lambda)$ by (10) we have to integrate ${\tilde{\chi}}(x,s)$ over 
the negative real halfaxis. Let us now move all the branch points shown on 
Fig.10b back according to moving back the point $x$ on the
contour $K'$ to 
its original position in the sector $S_1$. The point $s=0$ of this sheet is 
then left from the right. This motion causes the integration path just 
mentioned
 to be deformed into the one shown in Fig.2a. If we apply now to the 
cut emerging from the branch point $\xi-\zeta_1$ an operation of rotating it
clockwise by $\pi$ which is reversed to the uscreening operation described in 
Theorem 6 of Appendix 1 we obtain
 the situation drawn on Fig.2b. This figure shows in details 
why in these 
positions of the considered cuts the integration in (\ref{10}) along the path 
$C_3$ 
provides us with the factor $\chi_3(x,\lambda)$ which corresponds to the 
sector
 $S_3$ (i.e. the infinite integration limit in (\ref{5}) coincides with 
$\infty_3$), 
but is continued on the $x$-plane to the sector $S_1$ along the contour $K'$. 

Let us note further that in the positions of the cuts shown on
Fig.2b 
we can 
obtain the factors $\chi_1(x,\lambda)$ and $\chi_2(x,\lambda)$ as well 
integrating along the left and right real halfaxes.

\begin{tabular}{cc}
\psfig{figure=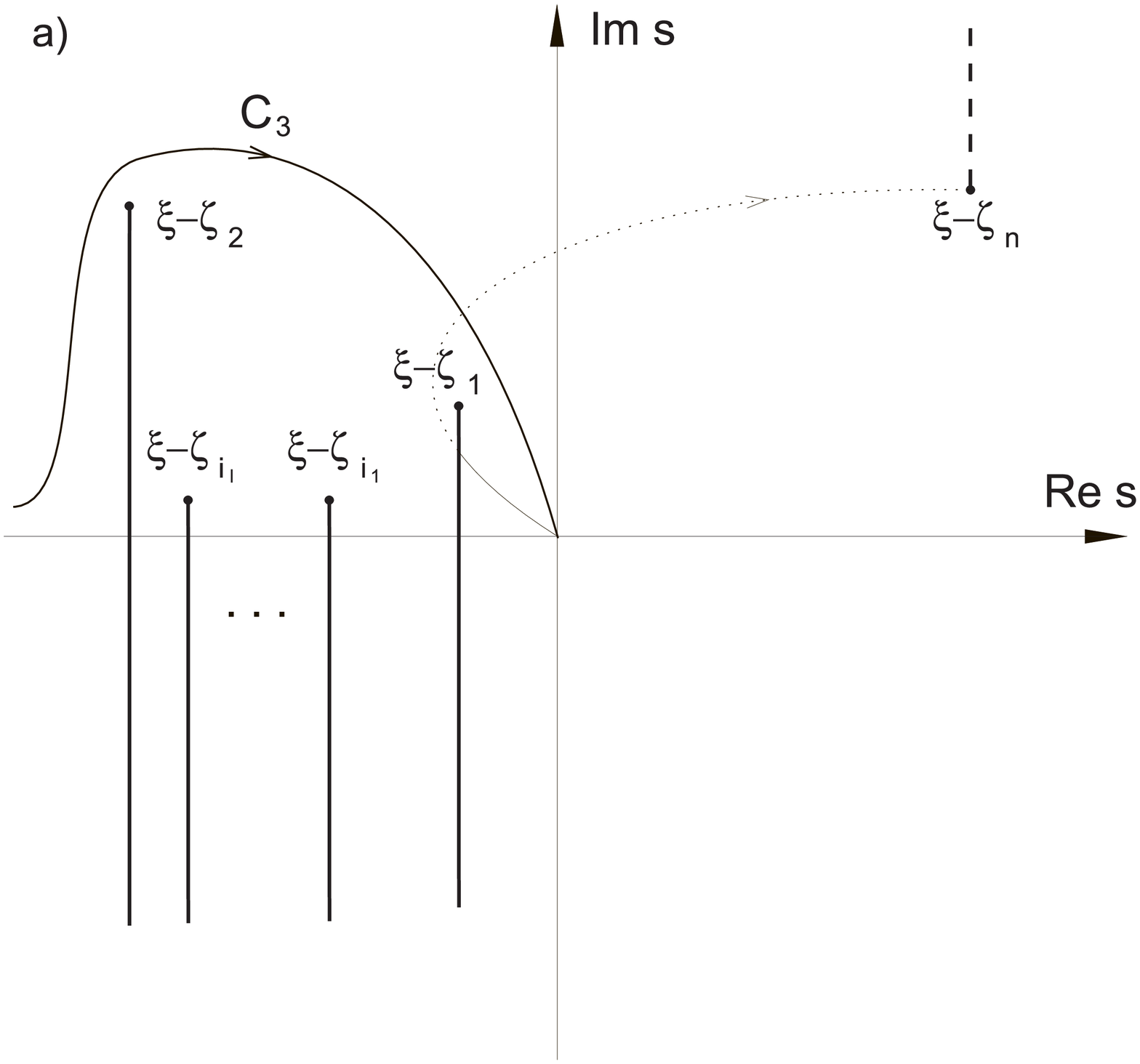,width=6.5cm} & \psfig{figure=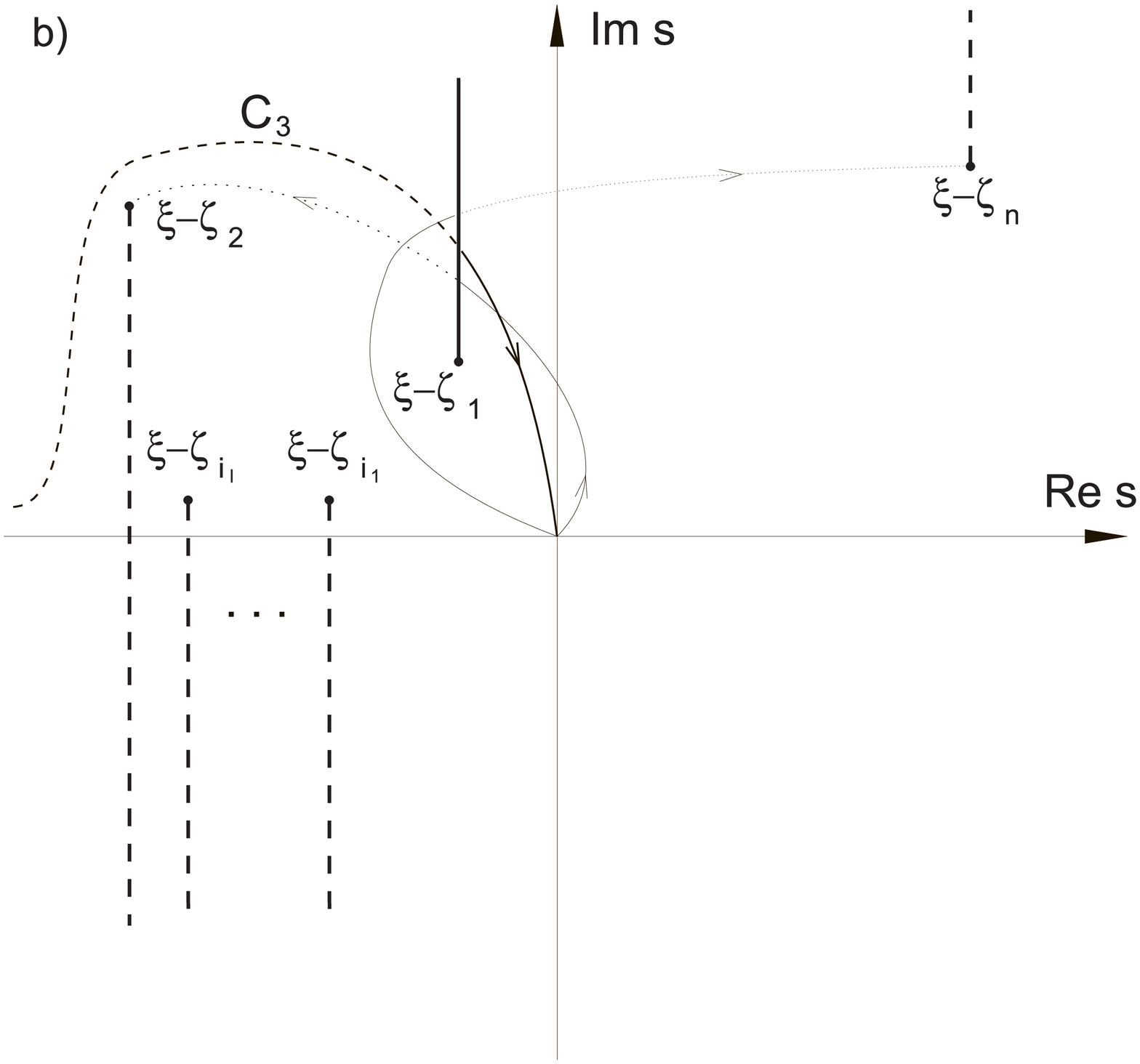,width=6.5cm} \\
Fig.2a The position of the branch points at &  Fig.2b Same as in
Fig.2a after uprighting\\
$\xi-\zeta_1,... ,\xi-\zeta_2$ and of the contour $C_3$&the cut
  emerging from the point $\xi-\zeta_1$\\
for $\tilde\chi(x,s)$ when $x\in K'\cap S_2$&$\;$
\end{tabular}

\vskip 15pt

We can repeat the above analysis starting from the cut pattern shown 
in Fig.11b
 and corresponding to the Borel function 
$\tilde\Phi(\xi(x),s)(\equiv\tilde\chi(x,s))$ continued to the sector $S_k$ 
along the contour $K'$. In this position of $\xi$ the factor 
$\chi_k(x,\lambda)$ is recovered by (\ref{10}) by integrating 
$\tilde\Phi(\xi,s)$ along, say, the negative real halfaxis (assuming $k$ is 
odd). Next, moving the point $x$ back to its original position in the sector 
$S_1$ and applying to the consecutive cuts emerging from the branch points at 
$\xi-\zeta_{i_{k-1}}$, $\xi-\zeta_{i_{k-2}}$,..., $\xi-\zeta_2$,
$\xi-\zeta_1$ the 
operations reversed to the uprighting ones described in Theorem 6 of 
Appendix 1
 we achieve the pattern of Fig.3. It follows from the figure that the above 
operations deform merely homotopically the integration contour $C_k$ from its 
original position when it coincides with the negative real halfaxis to the 
one 
shown in this figure where it has a spiral form allows it avoiding all the 
branch points.
 
\begin{tabular}{c}
\psfig{figure=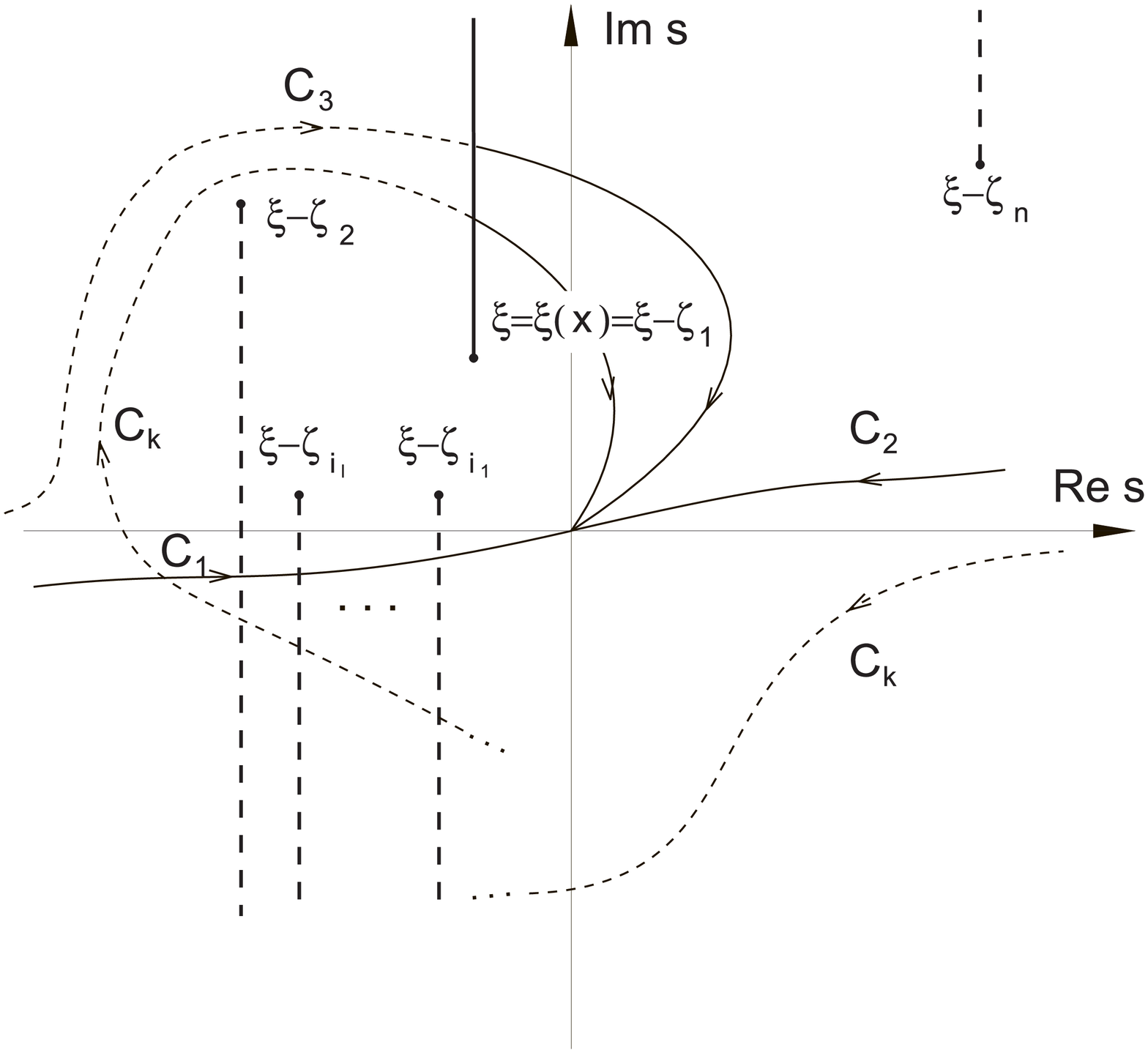,width=10cm} \\
Fig.3 The corresponding intgration paths $C_1,\;C_2,\;,...,\;C_k$ in the
  formula (10) \\when $x\in K'\cap S_1$
\end{tabular}

\vskip 15pt

Let us note that as it follows from the analysis performed in Appendix 1 
the $fixed$ branch points lie on the lower sheets of the $s$-Riemann surface 
beeing $always$
screened by the moving ones and therefore $not$ participating in the 
deformation of the contour $C_k$.

Note also that this deformation of the contour $C_k$ does not affect the 
convergence of the
integral (\ref{10}) since as we have shown in \cite{6} and mentioned in 
Appendix 1 on $each$ sheet of the $s$-Riemann surface the divergence of 
${\tilde{\Phi}}(\xi,s)$ at infinity is at most exponential.

In this way we have however completed the proof of the theorem. QED.

It is certainly worth to stress that a net result which follows from Theorem 1 
is that to obtain the subsequent 
${\chi}_{k}(x,\lambda)$'s, $k=2,3,...,n+2$ it is 
enough (according to our enumeration convention) to deform $C_1$ homotopically
anticlockwise by making its infinite tail to rotate by the angles 
$\pi, 2\pi,...,(n+1)\pi$ so that to coincide eventually with
the real halfaxes, positive or negative, on the corresponding 
sheets, see Fig.3. We get in this way
the sequence of paths $C_2$, $C_3$,...,$C_{n+2}$ integrations on which
according to the formula (\ref{10}) provide us with the corresponding
${\chi}_{k}(x,\lambda)$'s, $k=2,...,n+2$.  But these latter $\chi$-factors are
 exactly the ones which were obtained in Sec. II by applying to 
 ${\chi}_{1}(x,\lambda)$ the subsequent cyclic rotations since the latter 
correspond to (opposite) rotations on the $\lambda$-Riemann surface by the 
angles $-\pi,-2\pi,...,-(n+1)\pi$. It follows however from the formula 
(\ref{10})
that such cyclic rotations have to be accompanied then by the compensating
rotations of the integration path on the $s$-Riemann surface to maintain the 
convergence of the formula. 

We see therefore that the cyclic rotation property of transforming the 
fundamental solution $\psi_1(x,\lambda)$ into $\psi_k(x,\lambda)$ can be 
realized in the following two equivalent ways: 

$1^0$ on the $x$-plane 
by deforming the integration path $\gamma_1(x)$ in the formula (\ref{5}) 
(when it defines the solution $\psi_1(x,\lambda)$, with $x \in S_1$) into
the corresponding standard path linking the sector $S_1$ with the 
sector $S_k$; and

$2^0$ on the Borel plane by deforming the the path $C_1$ in the way 
described above.

We know however that after such $n+2$ cyclic rotations we do not come back to 
exactly the same factor ${\chi}_{1}(x,\lambda)$ but the latter acquires rather
 an additional phase factor which for the even degree polynomials is equal 
to $(-\imath)^n \exp(-\lambda \sigma_1 \oint_{K'} \sqrt{q(x,E)} dx)$. Therefore 
deforming the path $C_{n+2}$ once more in the above way to a path $C_{n+3}$ 
and integrating ${\tilde{\Phi}}(\xi,s)$ along this path we get as a result 
again ${\chi}_{1}(x,\lambda)$ but multiplied by the phase factor just 
mentioned i.e. for the even degree polynomials we have:
\be
\int_{C_{n+3}}e^{2\lambda s} {\tilde{\chi}}(x,s) ds =
e^{-\lambda \sigma_1\oint_{K'}\sqrt{q(y,E)}dy - \imath n\frac{\pi}{2}}
\int_{C_1}e^{2\lambda s} {\tilde{\chi}}(x,s) ds
\label{11}
\ee

Deforming $C_1$ appropriately clockwise we obtain of course the corresponding 
integration 
paths $C_{k \prime}^{\prime}$, $k^{\prime}=2,3,...$, providing us with 
${\chi}_{k}(x,\lambda)$'s ordered in the opposite way i.e.  with 
$k=n-k^{\prime}+4=n+2,n+1,...,2$. For the path $C_{n+3}^{\prime}$ we get an 
identity similar to (\ref{11}) but with the opposite sign at the exponent of 
the proportionality coefficient. This confirms that the $s$-Riemann surface 
of ${\tilde{\chi}}(x,s)$ is in general infinitely sheeted.
 The only obvious case with the finite six sheeted $s$-Riemann 
surface is provided by the linear potential \cite{6}.

Let us discuss still in some details the
deformation procedure of the path $C_1$ described above.  

The singularity pattern of Fig.3 which corresponds to $x \in S_1$  shows 
that to fall on the corresponding sheets in 
order to approach eventually the chosen direction of the real axis the paths 
$C_k$ have to avoid in general the existing singularities of 
${\tilde{\chi}}(x,s)$ on its $s$-Riemann surface. According to Fig.3 such
necessary deformations have to be applied for example to the path $C_3$ and 
to the subsequent ones but not to $C_2$.

The integration in (\ref{10}) along $C_3$ provides us with $\chi_3 
(x,\lambda)$ but since $x \in S_1$ the corresponding integration path
in (\ref{5}) cannot be then canonical i.e. for $\mid \lambda \mid \rightarrow
\infty$ $\psi_3 (x,\lambda)$ does not behave according to its JWKB factor in 
(\ref{3}). The obvious reason for that is just the (branch point) singularity 
of ${\tilde{\chi}}(x,s)$ at $s = \xi$ (with $\Re \xi > 0$) which causes 
$\chi_3 (x,\lambda)$ calculated in this way to diverge
as $e^{2\lambda \xi}$ in the semiclassical limit.

To restore, therefore, the proper canonical behaviour of $\chi_3 (x,\lambda)$
in this limit given by (\ref{8}) we would have to move the singularity at 
$s=\xi$ to the left halfplane of Fig.3 i.e. to move the corresponding 
variable $x$ from the sector $S_1$ to $S_2$ along the contour $K'$. This is 
just the procedure described in the course of the proof of the theorem.  

Earlier we have distinguished the canonical paths of integrations in 
(\ref{5}) as the ones which ensured that the $\chi$-factors in (\ref{3}) 
had well defined semiclassical expansions given by (\ref{8}). A part of 
these paths penetrating the domains $B_k$'s mentioned in the previous 
section ensured also that the fundamental solutions defined by them were 
Borel summable and these resummations were achieved by the Borel transforms 
of the Borel function $\tilde{\chi}(x,s)$ along halflines running from  
the infinity of the Borel 'plane' and ending at its center. Let us call 
$canonical$ also these latter paths on the Borel 'plane'. 

However we could notice above that it is possible to generalize substantially 
the notion of the Borel transformation by integrating in 
the Borel 'plane' along the paths $C_k$ described above and 
recovering in this way the fundamental solutions obtained by the 
deformations of the canonical paths in the formulae (\ref{5})
due to the cyclic rotations of the Stokes graph. 

It is therefore 
worthwhile to distinguish also these new types of the Borel transformation 
paths and these non canonical paths in the $x$-'plane' as well which appear 
as a result of the homotopic deformations of the canonical paths by the 
cyclic rotations. Namely, we shall call further such paths both on the 
$s$- and on the $x$-'plane' as the $standard$ paths in common.

Although it should be obvious that the Borel transformation of the Borel 
function $\tilde{\chi}(x,s)$ along any standard path should always provide 
us with the corresponding $\chi$-factor of the Dirac form (\ref{3}) of the 
fundamental solutions we show this fact explicitly in Appendix 2.

It is a good moment of our discussion to mention an old problem of the 
semiclassical theory known as the connection problem \cite{22,23,24,25}. 
In the context of our considerations it arises when 
we are interested in the semiclassical behaviour of $\psi_3 
(x,\lambda)$ whilst $x$ is kept in $S_1$ (i.e. $x \in S_1$).
In such a case we can 
deform the standard path $C_3$ into two paths, a path $C_{3}^{\prime}$ 
surrounding the
cut generated by the singularity at $s = \xi$ (see Fig.4) and again
the canonical path $C_1$. By multiplying (\ref{10}) (with $C_3$ as the 
integration 
path) by $q^{-1/4}e^{-\lambda \xi}$ we obtain $\psi_3 (x,\lambda)$ to be 
represented in this way by the following linear combination of two solutions 
to Schr\"odinger equation (\ref{2}): 
\be
\psi_3 (x,\lambda) = \psi_1 (x,\lambda) + C(\lambda) \psi_2 (x,\lambda)
\label{12}
\ee

\vskip 15pt

\begin{tabular}{c}
\psfig{figure=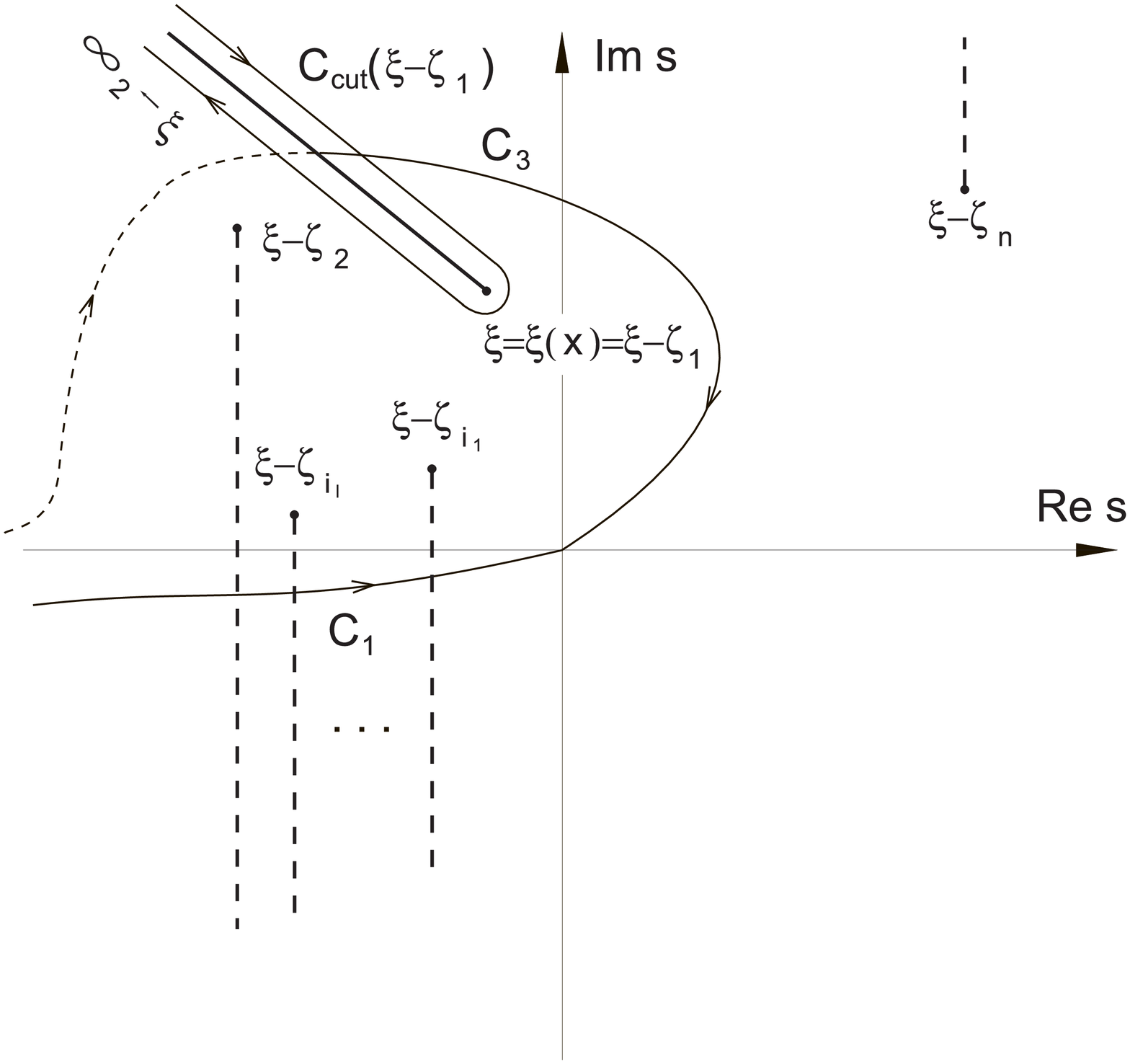,width=10cm} \\
Fig.4 The path $C_3$ splitted into the paths $C_1$ and
  $C_{cut}(\xi-\zeta_1)$
\end{tabular}

\vskip 15pt

Of course, $\psi_1 (x,\lambda)$ is generated by the $C_1$ part of $C_3$.
The fact that the cut integration part of $\psi_3 (x,\lambda)$ is just
proportional to $\psi_2 (x,\lambda)$ can be easily seen by pushing $\xi$ to
infinity along the cut what corresponds to approaching by $x(\xi)$
the infinity of the sector $S_2$. The cut integral (multiplied by
$q^{-1/4}e^{-\lambda \xi}$) vanishes however in this limit (since $\Re \xi
\rightarrow -\infty$) what proves our assertion. In other words we have: 
\be
q^{-\frac{1}{4}}(x,E) e^{- \lambda \int_{x_1}^{x}\sqrt{q(y,E)}dy}
2\lambda\int_{C_{cut}(\xi)}e^{2\lambda s} {\tilde{\chi}}(x,s) ds =
C(\lambda) \psi_2 (x,\lambda)
\label{13}
\ee

It is easily seen that the last relation is independent of such critical 
forms of the Stokes graph as the ones corresponding to coinciding of two 
(or more) turning points. However, when the Stokes graph is built 
only by the simple turning points then this relation can be established 
by the standard methods, i.e. by continuing the fundamental solutions along
 the canonical paths on the $x$-'plane', which give \cite{1,2}: 
\be
C(\lambda) = -\imath
\chi_{3\rightarrow 1}(\lambda)
\label{14}
\ee
where
$\chi_{3 \to 1}(\lambda) = \lim\limits_{x \to \infty_1}
\chi_3(x,\lambda)=\chi_{1 \to 3}(\lambda) $
is calculated by (\ref{5}) along the canonical path $\gamma_{1\to 3}$ 
(see Fig.1b).

Writing further $\psi_2(x,\lambda)$ in its Dirac's form (\ref{3}) we get:
\be
\chi_2(x,\lambda) = \imath \chi_{3\rightarrow 1}^{-1}(\lambda)
e^{- 2\lambda \xi}
2\lambda\int_{C_{cut}(\xi)}e^{2\lambda s} \tilde{\chi}(x,s) ds\\ =
\imath \chi_{3\rightarrow 1}^{-1}(\lambda)
2\lambda\int_{C _{cut}(0)}e^{2\lambda s} \tilde{\chi}(x,s+\xi) ds\nn
\label{L1}
\ee 

Taking now into account that $\chi_2(x,\lambda)$ is given by the Borel 
transformation along the canonical path $C_2$ on Fig.4 whilst 
$\chi_{3\rightarrow 1}(\lambda) (\equiv \chi_{1\rightarrow 3}(\lambda))$ 
can be obtained analogously by the integration 
${\tilde{\chi}}_{can}(\infty_3,s)(\equiv\lim\limits_{x\rightarrow\infty_3}\tilde{\chi}(x,s))$ 
along $C_1$ when $\xi(x)$ on Fig.1b goes from the sector $S_1$ to 
$\infty_3$ of the sector
 $S_3$ along the canonical path we get the following 
relation for ${\tilde{\chi}}(x,s)$ and its jump 
$\Delta_{s_0}\tilde{\chi}(x,s)$ through the cut emerging from 
the branch point $s=s_0=\xi(x)$:
\begin{eqnarray} 
\tilde{\chi}(x,s) = \imath {\tilde{\chi}_{can}}^{*-1}(\infty_3,s)*
\Delta_{0}\tilde{\chi}(x,-s+\xi(x))
\label{L2}
\end{eqnarray}

The '*' symbols in (\ref{L2}) denote the convolution operations (see
Appendix, formula (56)).

One easily recognizes in (\ref{L2}) the fundamental solution version of the 
analytical bootstrap property of the Borel function ${\tilde{\chi}}(x,s)$ 
discovered by Voros \cite{7}. It is the aim of this paper, however, to show 
that in the case of the polynomial potentials there are $no$ other versions 
of the realization of the analytical bootstrap idea since in this case the 
Borel function ${\tilde{\chi}}(x,s)$ is unique (up to an irrelevant constant) 
being uniquely defined by the fundamental solutions.

The same comments as above are valid of course with
respect to the results of the integrations along the subsequent
standard paths $C_k$, $k=4,5,...$, i.e. these paths provide us with 
the corresponding
$\chi_k (x,\lambda)$'s calculated along the non canonical standard paths 
on the $x$-'plane' obtained by the continuation of the variable $x$ from 
$S_k$ to $S_1$ along the contour $K'$. Such a form 
does not allows us for estimating easily its semiclassical limit. To 
recover this limit properly we have to deform $C_k$'s keeping its infinite 
tail along the appropriate real halfaxis. This deformation 
splits $C_k$ into $C_1$ or $C_2$ (the latter choice depends on a sign 
of $\Re \lambda$) and into a number of paths surrounding some cuts 
(the cuts have 
to run to the left halplanes for $\Re \lambda >0$ or to the right ones 
in the opposite case). Each such a cut contribution represents a solution 
to Schr\"odinger equation (see Appendix 2) being
proportional to some fundamental solution. The identification of these 
solutions can be performed by considering the limit of the latter when 
$\xi \rightarrow \infty$ along the appropriate cuts (or their elongations) 
(the solutions have to vanish in this 
limit) and following parallelly the corresponding paths drawn by $x(\xi)$ on 
the Stokes graph. 
Again, if the Stokes graph considered is determined only by simple turning 
points then a total number of the 
fundamental solutions engaged in the above splitting operation is limited 
only to those of them which can contact canonically with the sector 
$S_1$ where 
the variable $x/\xi$ ($\xi=\xi(x)$) actually is and the proportionality 
coefficients of the cut contributions to appropriate fundamental solutions 
can be calculated by the standard methods \cite{1,2}. 

Let $x$ be fixed somewhere on $K'$ (see Fig.1a).
 
We shall call a cut path each path surrounding a halfline cut of the 
$s$-Riemann
surface runnig from its infinity and ending at some of its moving 
or fixed branch points.  

Together with the result of Appendix 2 the net results of the
above discussion can be summarize as 
the following two theorems.

\begin{moje}
a) The Borel function $\tilde{\chi} (x,s)$ when Borel transformed along a
standard or a cut path and multiplied by the JWKB factor always
provides us with a solution to the Schr\"odinger equation $(\ref{2})$
having the Dirac form $(\ref{3})$; 

b) The solutions we get in this way
are always proportional to fundamental solutions;

c) If a solution defined by a cut path is generated by the deformation 
of the standard path then if it is proportional 
to the fundamental solution $\psi_k(x,\lambda)$ with 
a proportionality constant $C_k(\lambda)$ then a jump 
$\Delta_{s_0}\tilde{\Phi}(\xi(x),s)$ of $\tilde{\Phi}(\xi(x),s)(\equiv 
\tilde{\chi}(x,s))$ through the cut 
generated by the branch point at $s=s_0$ is 
related in the following way with $\tilde{\Phi}(\xi,s)$ itself (a case 
of the analytic bootstrap of Voros $ \cite{7}$):
\begin{eqnarray}
\tilde{\Phi}((-1)^{k-1}\xi(x'),s) = 
\tilde{C}_k(s,s_0)*\Delta_{s_0}\tilde{\Phi}(\xi,\pm s+s_0)
\label{L3}
\end{eqnarray}
where the RHS of $(\ref{L3})$ is the convolution of  $\tilde{C}_k(s,s_0)$ 
which is the Borel function corresponding to the 
invers of the constant $C_k(\lambda)$ multiplied by 
$e^{2\lambda s_0}$ and of 
$\tilde{\Phi}(\xi,s)$ shifted respectively 
whilst $\tilde{\Phi}((-1)^{k-1}\xi(x'),s)$ 
denotes $\tilde{\chi}(x,s)$ continued analytically along $K'$ from the point 
$x\in K'\cap S_1$ to the point $x'\in K'\cap S_k$ such that 
$\xi(x)=(-1)^{k-1}\xi(x')$. The '$\pm$' signs at the variable $s$ in $(\ref{L3})$
takes into account that Borel integrations along the cut and the 
canonical path $C_k^{can}$ to recover $\chi_k(x,\lambda)$ can go to the 
same ('$+$' sign) or to the opposite ('$-$' sign) infinities of $\Re s$.

The existence of $\tilde{C}_k(s,s_0)$ and the holomorphicity of 
$\Delta_{s_0}\tilde{\Phi}(\xi(x),\pm s+s_0)$ at $s=0$ is assumed.

\end{moje}

\vskip 6pt
{\it Proof of the theorem} 

The part a) of the theorem is obvious by
noticing that any Borel transformation along the standard/cut path
with the Borel function $\tilde{\chi} (x,\lambda)$ defined by (\ref{9}) 
satisfies the linear
differential equation defining $\chi (x,\lambda)$'s (see Appendix 2).  

The part b) of the theorem when concerning the standard paths is 
a repetition of the corresponding result of Theorem 1. With respect, 
however, to the cut paths it follows as a conclusion summarizing the 
discussion preceding the formulation of this theorem.

The part c) is the direct consequence of the hypothesis of this part of the 
theorem written explicitly as:
\begin{eqnarray}
q^{-\frac{1}{4}}(x')e^{\sigma_k\lambda\xi(x')}
2\lambda\int_{C_{cut}(s_0)}e^{2\sigma\lambda s}
\tilde\Phi(\xi(x),s)ds = C(\lambda)q^{-\frac{1}{4}}(x')
e^{\sigma_k\lambda \xi(x')}
\chi_k(x',\lambda)\\ = C(\lambda)q^{-\frac{1}{4}}(x')e^{\sigma_k\lambda \xi(x')} 
2\lambda\int_{C_k^{can}}e^{-2\sigma_k\lambda s}
\tilde\Phi(\xi,s)ds\nn
\label{L4}
\end{eqnarray}
where $C_k^{can}$ denotes the canonical path on the Borel 'plane' used 
for recovering $\chi_k(x,\lambda)$, $\sigma_k$ is a signature of the latter 
and $\sigma=\pm1$ is taken as $+1$ for the integration along the cut 
$C_{cut}(s_0)$ to $-\infty$ of $\Re s$ and as $-1$ in the opposite case.

Since $\chi_k(x',\lambda)$ in (\ref{L4}) is recovered by the integration 
$\tilde\Phi(\xi(x'),s)$ along the $canonical$ path $C_k^{can}$ then it 
means that this 
latter function results as its analytical continuation from the first sheet 
where it determines $\chi_1(x,\lambda)$ (by the integration along 
a canonical path $C_1$ on this sheet) to the sheet considered. 
This continuation 
has to be performed along the path $\tilde K'$, the image of $K'$ on the 
Borel 'plane', as long as $\xi$ acquires the argument equal to $(k-1)\pi$ 
which put $\xi$ in the position $(-1)^{k-1}\xi$ on the final sheet. This 
corresponds of course to a point $x'$ on $K'$ such that 
$\xi(x)=(-1)^{k-1}\xi(x')$. 

Now, if $\sigma_k=\sigma$ (this case is the only possible when $s_0$ is the 
fixed branch point) then we get the relation (\ref{L3}) with '$+$' at 
the variable $s$ and with '$-$' in the opposite case.  

This latter conclusion ends, however, the proof of the theorem. QED.

Let us note, as a comment to the part (b) of the above theorem, that the 
proportionality constants can be always calculated independently when all 
turning points of considered polynomial problems are simple. We can use 
then the powerful method of analytical continuation of the fundamental 
solutions along the canonical paths which guarantees the full controll 
of the semiclassical properties of calculated quantities at each stage 
of such calculations. The considerations preceding the above theorem are 
the good illustration of the possibilities of the method.

\begin{moje} 
The connection problem i.e. the analytical
continuation of the fundamental solutions throughout the x-plane
along the contour $K'$ of Fig.1a can be solved by performing this
continuation on the Borel plane. By such a continuation the
original Borel integration along the deformed path has to be
splitted into integrations along standard and cut paths the
latter emerging from the branch points of $\tilde{\chi}(x,s)$ pinching the
deformed path.
\end{moje}

\vskip 12pt

$Proof$. The validity of the theorem follows directly from the
preceding discussion.

\vskip 6pt

Another important property of the fundamental solutions which
distinguishes these solutions among other possible Borel
summable solutions can be formulated as the following theorem.

\begin{moje}  
Let $x_0$ ($=x(\xi_0)$) be an $arbitrary$ point of the $x$-plane
not coinciding with a root of $q(x,E)$. Then there is a (non
empty) subset $N(x_0)$ of fundamental solutions of both signatures
with the following properties: 

$1^0$ The point $x_0$ is canonical for every member of $N(x_0)$;

$2^0$ Every element of $N(x_0)$ can be obtained by the formula (\ref{10}) 
integrating along a corresponding standard path.
\end{moje}

\vskip 12pt

We shall assume $N(x_0)$ to collect all such fundamental solutions.

{\it Proof of the theorem}

The validity of this theorem can be easily seen by considering the
topology of sectors with respect to the chosen $x$ on the Riemann
surface of the action variable $\xi (\equiv \xi(x))$ substituting the
variable $x$ (see Fig.5 and \cite{1}). For real $\lambda$ the Stokes lines 
on the surface are now parallel to imaginary axes and the sectors
are left and right halfplanes not containing (the images of)
turning points on each sheet of the surface \cite{1}. The $\lambda$-rotations
of Stokes graph make Stokes lines on the $\xi$-Riemann surface
rotating around the images of the turning points preserving their
parallelness.

\vskip 15pt

\begin{tabular}{cc}
\psfig{figure=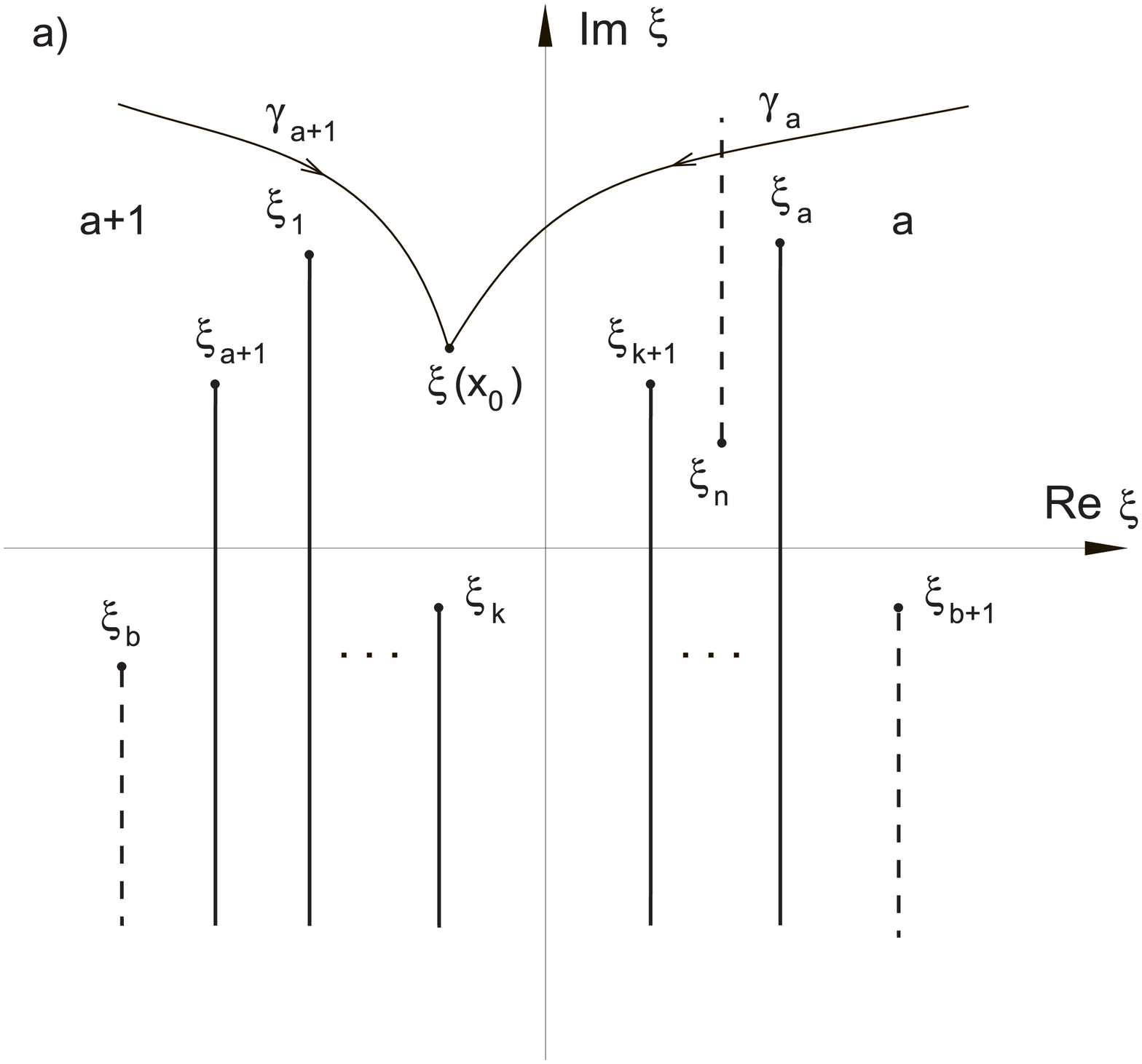,width=6.5cm} & \psfig{figure=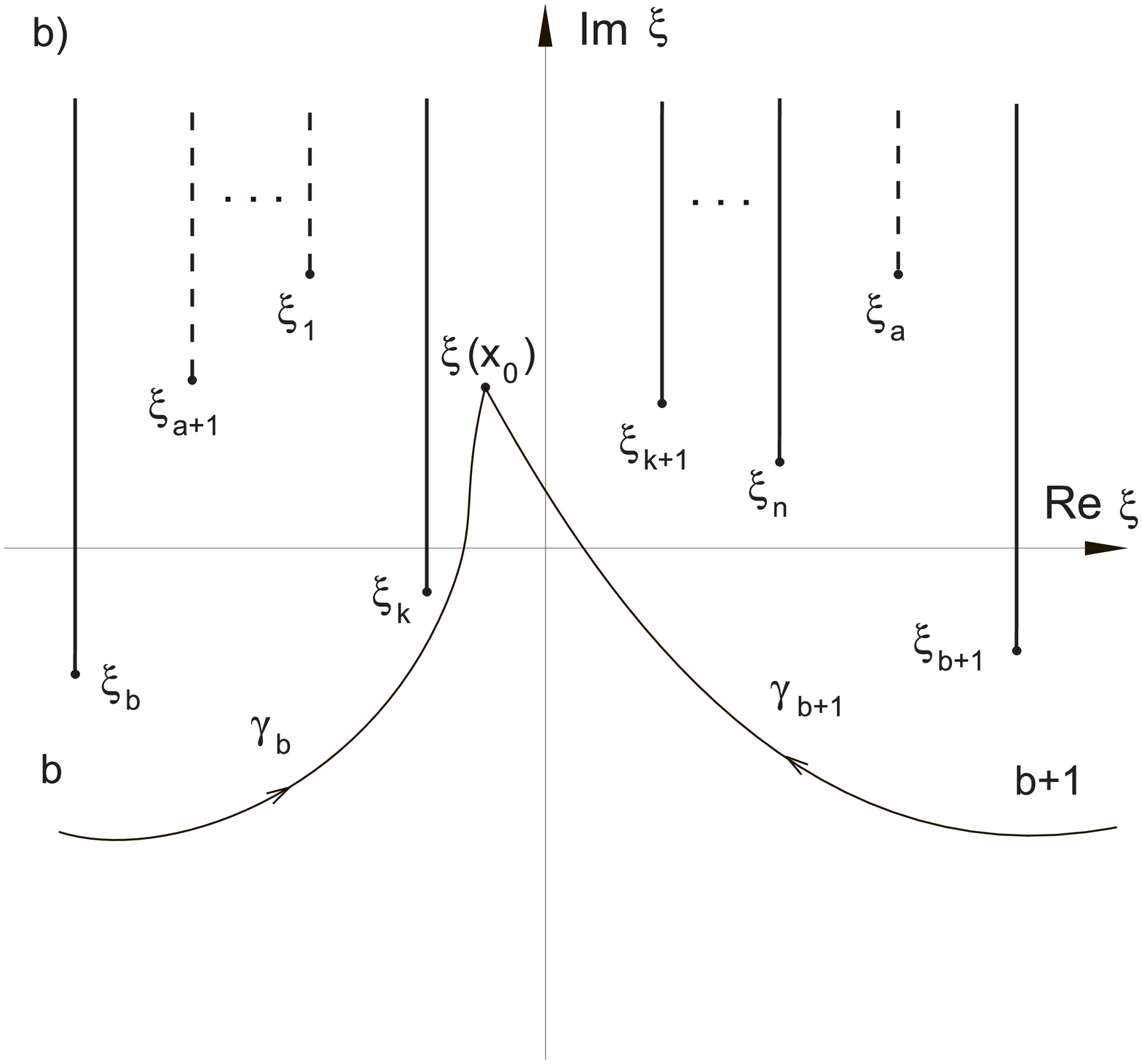,width=6.5cm} \\
Fig.5a The sectors $S_{a+1}$ and $S_a$ the funda-&  Fig.5b The
sectors $S_b$ and $S_{b+1}$ the funda-\\mental solutions of which
communicate&mental solutions of which communicate\\canonically with the point $\xi(x_0)$&canonically with the $same$ point $\xi(x_0)$
\end{tabular}

\vskip 15pt

We shall distinguish the following four cases for the position of $x_0$ for 
real $\lambda$:
\begin{enumerate}
\item $x_0$ does not lie on any Stokes line but
\begin{description}
\item[a.] $x_0$ belongs to some sector,

\item[b.] $x_0$ does not belong to any sector;
\end{description}
\item $x_0$ lies on some Stokes line and
\begin{description}
\item[c.] this Stokes line is finite i.e. it emerges from some turning point and ends
on the other,

\item[d.] this Stokes line is infinite i.e. it emerges from some turning point and
runs to infinity of the $x$($\xi$)-plane.
\end{description}
\end{enumerate}

The above possibilities exhaust of course all the possible positions of $x_0$
with respect to the Stokes lines.

Consider the case $1^0${\bf a.}

First let us note that for real $\lambda$ the Stokes lines are parallel to 
the imaginary axis as it is shown on Fig.5. If $x_0$ is in some sector, say  
$S_a$, then obviously the fundamental solution $\psi_a$ defined in this 
sector belongs to $N(x_0)$. There are however another two fundamental 
solutions which belong to $N(x_0)$ too. Namely, these are the two neighbours 
of $\psi_a$, i.e. $\psi_{a+1}$ and $\psi_{a-1}$ (according to our enumeration 
convention). Both of them
have their signatures opposite to $\psi_a$. Depending on a position of $x_0$
in $S_a$ there is always one of these two neighbour solutions which is Borel 
summable at $x_0$ simultaneously with $\psi_a$ and at the same time can 
communicate with it canonically. 

Consider next the case $1^0${\bf b.}

In this case $x_0$ is in some infinite vertical strip on the $\xi$-Riemann 
surface (see Fig.5) bounded from each side by two chains 
of Stokes lines parallel to each other. Following each of these chains of 
lines whilst 
keeping on the strip we find on them their extremal turning points i.e. the 
ones from which Stokes lines emerge running to imaginary infinities, positive 
or negative. Consider such a Stokes line which bounds the strip from the 
right and runs to positive imaginary infinity. This Stokes line bounds 
simultaneously from the left a strip neighbour to the one just considered. 
This new strip is again bounded from the right by a chain of Stokes lines 
which structure is similar to the Stokes line chains already considered. 
So there is again in this chain a Stokes line ending it and running to 
positive imaginary infinity. We can repeat this procedure of moving to 
the right of the $\xi$-'plane' to finish eventually with a 'strip' 
which is a right halfplane of the $\xi$-plane sheet considered. 
This procedure has to be finite since there is a finite number of extremal 
turning points met in this way due to the polynomial potential. It is clear
also
that this halfplane corresponds to a sector, say $S_a$, of the Stokes graph.
The way of its finding proves that the fundamental solution $\psi_a$ 
defined in it belongs to the set $N(x_0)$. Mutatis mutandis to this set
belongs also the solution $\psi_{a+1}$ corresponding to the sector $S_{a+1}$,
the next one to $S_a$ and having an opposite signature. By the same resonings
 but moving down the $\xi$-sheet 
we find two others sectors, call them $S_b$ and $S_{b+1}$, the fundamental 
solutions of which are both Borel summable at $x_0$ having of course opposite 
signatures. Therefore there are at least four fundamental solutions in 
$N(x_0)$ for this case. Besides, within the pairs $\psi_a, \psi_b$ and 
$\psi_{a+1}, \psi_{b+1}$ both the fundamental solutions can communicate with 
themselves canonically and can be simultaneously Borel summable. 
This last result is a conclusion from the fact that 
these pairs of fundamental solutions are related by the cyclic rotations.

Consider now the case $2^0${\bf c.}

In this case we can consider both the sides of the Stokes line on which $x_0$ 
is placed as pieces of two strips for which this line is their bound. The 
strips can be identified as the wholes by the procedure similar to the one used 
previously. Namely, we move along a chain of Stokes lines directing to, say, 
the positive imaginary infinity choosing the most right Stokes line each time
we meet some turning point. Continuing this motion we meet finally the last 
such turning point from which the most right Stokes line emerges running 
directly to the positive imaginary infinity. From this moment we can repeat 
the arguments of the previous point when concluding that we can find in the 
right halfplane on a certain $\xi$-sheet a sector, call it $S_a$ again, a 
fundamental solution of which is Borel summable at $x_0$ i.e. this solution 
belongs to $N(x_0)$. 

However, this case differ a little bit from the previous 
one by the fact that the neighbour sector $\psi_{a+1}$ is $not$ Borel summable 
at $x_0$ in the actual position of this variable. Nevertheless, if we rotate 
the variable $\lambda$ anticlockwise by an arbitrary small angle we 
immediately satisfy the Watson-Sokal-Nevanlinna conditions for the Borel 
resummation of $\psi_{a+1}$. Making a cyclic rotation in the same direction 
we transform the last solution into $\psi_a$ and this rotation can still be
continued a little bit further. This proves that $\psi_a$ indeed satisfies
the conditions mentioned above to be Borel summable at $x_0$ in its actual 
position. 

Again mutatis mutandis we can prove moving down the $\xi$-sheet 
the existence of another fundamental solution in the right half of the sheet 
which is Borel summable at $x_0$. Let us call it $\psi_b$. 

Repeating the procedure and keeping the most left Stokes lines whilst 
moving to the imaginary infinities in both directions, positive and negative, 
we find still another two respective solutions $\psi_d$ and $\psi_c$. 
Collecting them into the following pairs $\psi_a, \psi_c$ and $\psi_b, 
\psi_d$ we find that the solutions in each pair communicate canonically with 
each other and are Borel summable simultaneously at $x_0$.

In this way we have proved however our theorem. QED.

Theorem 4 is, to some extend, a generalization of Theorem 1. 
Namely, we have:

{\it\bf Corollary}

{\it In 
the assertion (b) of Theorem 1 we can take $any$ regular point of the 
$x$-plane in which we want to obtain $any$ fundamental solution continued 
to this point by the $cyclic$ rotation operations defined in Section II.} 

{\it Proof}

Our reasoning is the following. 

Theorem 4 tells us that for a given (regular) $x_0$ there is $always$ a 
fundamental solution, say $\psi_a(x,\lambda)$, Borel summable at this point. 
This means however that the $\chi$-factor of this solution is recovered from 
the Borel function $\tilde{\chi}(x,s)$ by the Borel transformation along the 
path $C_a$ coinciding with the left/right real halfaxis of the Borel plane. 
The latter plane is of course a sheet of the $s$-Riemann surface obtained 
from the one on Fig.7 by the unscreening operation. Depending on the 
actual position of $x_0$ the representation of branch points on this sheet 
can be richer than in the cases considered in Theorem 1 because of the closest
environment of $x_0$ which can be richer in turning points. This structure 
can be analyse in a way similar to that in Appendix 1 giving typical branch 
point pattern with both moving and 
fixed branch points on this sheet lying above or below the path $C_a$
but allowing to perfom the Borel transformation along the path.

We can now start to deform homotopically the path $C_a$ exactly in the same 
way as we
did with the path $C_1$ in Theorem 1 taking its infinite end and rotating it
by the angles $\pm\pi, \pm 2\pi,...$, to appropriate
positions along the left/right halfaxis on the sheets corresponding to 
subsequent fundamental solutions starting from $\psi_{a+1}(x,\lambda)$ in the
clockwise direction of the deformation or from $\psi_{a-1}(x,\lambda)$ in the
opposite case. This procedure could be disturbed only when the deformed 
path $C_a$ met on its way a chain of branch points elongating to an infinity. 
This is $not$ possible however since such possible chains of branch points are
always screened by the moving cuts. A good illustration of the described 
situation is provided by the harmonic oscillator case (see \cite{6}).

We have already mentioned 
in the course of the proof of the Theorem 1 that the suitable deformations of 
the path $C_1$ to $C_2,...,C_{n+2}$ described in this theorem to restore all 
the fundamental 
solutions when $x \in S_1$ strictly corresponded to the recovering these 
solutions from $\psi_1(x,\lambda)$ defined in $S_1$ by the cyclic rotations. 
Exactly the same relation connects the above deformations of the path $C_a$ 
with the cyclic rotations of the Stokes graph from its position just 
considered. 

The last conclusion finishes the proof of Corollary. QED.

\section*{IV.  General form of semiclassical expansion for $\chi$-factors}

\hskip+2em Let us note that the $\chi$-factors entering the Dirac forms (\ref{3}) are 
the solutions of the following two second order linear differential
equations obtained by the substitution (\ref{3}) into the 
Schr\"odinger equation: 
\begin{eqnarray}
- q^{- \frac{1}{4}}(x) \left( {q^{- \frac{1}{4}}(x)} \chi(x) 
\right)^{\prime{\prime}}
+2\sigma \lambda \chi^{\prime}(x) = 0
\label{15}
\end{eqnarray}

The equations (\ref{15}) provide us with a general form of
semiclassical expansions for the $\chi$-factors if such expansions exists. 
Namely,  assuming the latter we can substitute into (\ref{15}) 
the semiclassical expansion for $\chi$: 
\begin{eqnarray}
\chi (x,\lambda) \sim \sum_{n\geq{0}} 
\left( \frac{\sigma}{2\lambda} \right)^{n} \chi_{n}(x)
\label{16}
\end{eqnarray}
to get the following recurrent relations for $\chi_n (x)$: 
\begin{eqnarray}
\chi_n (x) = C_n +
\int_{x_{n}}^{x} q^{-\frac{1}{4}}(y) 
\left( q^{-\frac{1}{4}}(y) \chi_{n-1}(y) \right)^{\prime{\prime}} dy  
& \mbox{  ,  } \mbox{} n \geq 1 
\label{17}
\end{eqnarray}
\begin{eqnarray*}
\chi_0 (x) \equiv C_0 
\end{eqnarray*}
where $x_n$, $n \geq 1$, are
arbitrary chosen regular points of $\omega (x)$ and $C_n$, $n \geq 0$, are
arbitrary constants. It is, however, easy to show that choosing
all the points $x_n$ to be the same, say $x_0$, merely redefines the
constants $C_n$. Assuming this we get for $\chi_n (x)$:
\begin{eqnarray*}
\chi_n (x) = \sum_{k=0}^{n} C_{n-k} I_k (x,x_0)
\end{eqnarray*}
\begin{eqnarray*}
I_0 (x,x_0) \equiv  1
\end{eqnarray*}
\begin{eqnarray}
I_k (x,x_0) =
\int_{x_0}^{x}d\xi_{k}q^{-\frac{1}{4}}(\xi_{k}) 
\left( q^{-\frac{1}{4}}(\xi_{k}) 
\int_{x_0}^{\xi_{k}}d\xi_{k-1}q^{-\frac{1}{4}}(\xi_{k-1}) 
\left( q^{-\frac{1}{4}}(\xi_{k-1}) \int_{x_0}^{\xi_{k-2}}d\xi_{k-2} \ldots 
\right. \right.       
\label{18}
\end{eqnarray}
\begin{eqnarray*}
\left. \left.  
\times \left( q^{-\frac{1}{4}}(\xi_{2}) 
\int_{x_0}^{\xi_{2}}d\xi_{1}q^{-\frac{1}{4}}(\xi_{1}) 
\left( q^{-\frac{1}{4}}(\xi_{1}) \right)^{\prime{\prime}} 
\right)^{\prime{\prime}} \ldots 
\right)^{\prime{\prime}} \right)^{\prime{\prime}} 
\end{eqnarray*}
\begin{eqnarray*}
k = 1,2,\ldots 
\end{eqnarray*}

Substituting (\ref{18}) into (\ref{16}) we get finally for the expansion: 
\begin{eqnarray}
\chi (x,\lambda) \sim \sum_{n\geq{0}} 
\left(\frac{\sigma}{2\lambda} \right)^{n} C_n
\sum_{k\geq{0}} 
\left(\frac{\sigma}{2\lambda} \right)^{k} I_k (x,x_0)
\label{19}
\end{eqnarray}

In this way we have proven the following lemma

\vskip 18pt

{\it\bf Lemma 1 }

{\it An arbitrary semiclassical expansion $(\ref{16})$ which follows
from $(\ref{15})$ can be given the form $(\ref{19})$ with an arbitrarily chosen
regular point $x_0$ and arbitrary constants $C_n$, $n\geq 0$.  }

\vskip 12pt
We shall call (\ref{19}) the standard form of the expansion (\ref{16}).  

Of course, for a given $\chi$ the choice of $x_0$ determines the constants
i.e. the latter depend on it. However, if such a $\chi$ is given a
choice of $x_0$ cannot be arbitrary. The reasons for that are that
if $\chi$ considered can be semiclassically expanded then a domain of
the $x$-plane for such an expansion is strictly determined. Good
examples of the latter statement are provided just by the
fundamental solutions. Each of the latter possesses as we have
discussed it in Sec. 2 its allowed canonical domain of the
semiclassical expansion (\ref{19}). Therefore each $\chi$ possesses its own
domain $D_{\chi}$ of the existence of the corresponding semiclassical
expansion $\chi^{as}$. Such a domain can however be also empty (see
below).  

Suppose $D_{\chi}$ to be not empty and let $x, x_0 \in D_{\chi}$. 
Then we can expand $\chi$ semiclassically and 
this expansion has the form (\ref{19}).
Let us assume a little bit more about $\chi$, namely that there is a
domain $B_{\chi} \subset D_{\chi}$ in which $\chi$ is Borel 
summable and let $x, x_0 \in B_{\chi}$. Then
both $\chi(x,\lambda)$ and $\chi(x_0,\lambda)$ can be restored by the Borel
transformation of the corresponding Borel functions and,
respectively, along the negative real halfaxis (by assumption)
of the Borel plane. Their semiclassical expansions (\ref{19}) can be
obtained then by substituting simply into the Borel integral the
Borel series (9) with the respective arguments $x$ and $x_0$. But it
means, of course, that we can obtain $\chi^{as}(x_0,\lambda)$ simply from
$\chi^{as}(x,\lambda)$ by putting $x=x_0$ in the latter. Doing this in 
(\ref{19}) we see that it takes in this case the following form
\begin{eqnarray}
\chi (x,\lambda) \sim \chi^{as} (x,\lambda) = \chi^{as} (x_0,\lambda)
\sum_{k\geq{0}} 
\left(\frac{\sigma}{2\lambda} \right)^{k} I_k (x,x_0)
\label{20}
\end{eqnarray}

Therefore the following lemma has been proven

\vskip 18pt
{\it\bf Lemma 2}

{\it If $\psi(x,\lambda)$ is a solution to the Schr\"odinger equation 
$(\ref{2})$ given in some domain B in the Dirac form $(\ref{3})$ with the
corresponding factor $\chi(x,\lambda)$ having in B the standard
semiclassical expansion $(\ref{19})$ which is Borel summable in B to the
factor $\chi(x,\lambda)$ itself then this semiclassical expansions takes in
B the form $(\ref{20})$ where $x_0 \in B$.}
\vskip 12pt

The above formula shows explicitly the way of determining the
series of the constants $C_n$ in the case just discussed. However,
we shall show below that in general the form (\ref{20}) can not be
valid i.e. the series of constants in (\ref{19}) {\it is not} a
semiclassical expansion of $\chi(x,\lambda)$ at $x=x_0$ even if the
corresponding semiclassical expansions exist in both of the points.  

Nevertheless, the formula (\ref{20}) can be certainly applied
to the fundamental solution $\chi$- factors $\chi_k(x,\lambda)$ with 
$\chi_k^{as}(x,\lambda)$ and $\chi_k^{as}(x_0,\lambda)$ defined by (\ref{8})
when $x,x_0 \in B_k \subset D_k$, with $D_k$ being the
canonical domain of $\chi_k(x,\lambda)$. In these latter cases the formula
(\ref{20}) can be derived directly from (\ref{8}) by noticing that
\begin{eqnarray}
\chi_{k,n}(x) =
\sum_{p=0}^n  \chi_{k,p}(x_0) I_{n-p} (x,x_0)
\label{21}
\end{eqnarray}
and by multiplying both sides of (\ref{21}) by $(2\sigma \lambda)^{-n}$ and
summing over $n$ (from 0 to $\infty$).

\section*{V.  Other solutions with well defined
Borel summable semiclassical asymptotics}

\hskip+2em In this section we shall show that at
each point of the $x$-plane not coinciding with the root of $q(x,E)$
there are two pairs of base solutions to (\ref{15}) each of which can
be expanded semiclassically in some well defined domain. These
expansions are Borel summable in corresponding domains although
not to the solutions themselves.

\vspace{12pt}
\hskip+2em{\bf 1. Fr\"oman and Fr\"oman construction of solutions to
  Schr\"odinger equation }

A construction of the solutions just mentioned is the
following (see App. B in \cite{10} and \cite{3}).

In the $x$-plane we choose any point $x_0$ (being not a root of $q(x)$
however). The point distinguishes a line $\Re W_k(x,E) = \Re W_k(x_0,E)$ 
(it is independent of $k=1,2,...,n$) on which it lies so that $\Re W_k(x,E)$
increases on one side of the line and decreases on the other. On each 
side of the line we can define two independent solutions each having the 
form (\ref{3}) with the following formulae for the $\chi$-factors
(see App. B in \cite{10} and \cite{3}):
\begin{eqnarray}
\chi_{1}^{\sigma}(x,x_0) = 1 + \sum_{n{\geq}1}
\left( \frac{\sigma}{2\lambda} \right)^{n} \int_{x_0}^{x}d{\xi_{1}}
\int_{x_0}^{\xi_{1}}d{\xi_{2}} \ldots 
\int_{x_0}^{\xi_{n-1}}d{\xi_{n}}
\omega(\xi_{1})\omega(\xi_{2}) \ldots \omega(\xi_{n}) 
\label{22} \\
{\times} \left( 1 -
e^{-2\sigma{\lambda} {(W_{k}(x)-W_{k}(\xi_{1}))}} \right)
\left(1 - e^{-2\sigma{\lambda} {(W_{k}(\xi_{1})-W_{k}(\xi_{2}))}} 
\right)
\cdots \left(1 - e^{-2\sigma{\lambda} 
{(W_{k}(\xi_{n-1})-W_{k}(\xi_{n}))}} 
\right) \nonumber
\end{eqnarray}
and 
\begin{eqnarray*}
\chi_{2}^{\sigma}(x,x_0) = 
\frac{\sigma}{2\lambda}\frac{1}{q^{\frac{1}{2}}(x_0)}
\left[ 1-e^{-2\sigma{\lambda} (W_{k}(x)-W_{k}(x_0))}
+ \sum_{n{\geq}1}
\left( \frac{\sigma}{2\lambda} \right)^{n} \int_{x_0}^{x}d{\xi_{1}}
\int_{x_0}^{\xi_{1}}d{\xi_{2}} \ldots \right. 
\end{eqnarray*}
\begin{eqnarray}
\times
\int_{x_0}^{\xi_{n-1}}d{\xi_{n}}
\omega(\xi_{1})\omega(\xi_{2}) \ldots \omega(\xi_{n}) 
\label{23}
\end{eqnarray}
\begin{eqnarray*}
\left.
\times
\left( 1-e^{-2\sigma{\lambda} (W_{k}(x)-W_{k}(\xi_{1}))} \right)
\left( 1-e^{-2\sigma{\lambda} (W_{k}(\xi_{1})-W_{k}(\xi_{2}))} \right)
\ldots 
\left( 1-e^{-2\sigma{\lambda} (W_{k}(\xi_{n})-W_{k}(x_0))} \right)
\right] 
\end{eqnarray*}
where $\sigma = +1$ for $x$ on the side of
increasing $\Re W_k (x,E)$ and $\sigma =-1$ in the opposite case so that all
integrations in (\ref{22}) and (\ref{23}) run from $x_0$ to $x$ along the
canonical paths, finite this time. Due to that both the solutions to 
Schr\"odinger equation obtained by multiplying the $\chi$-factors (\ref{22}) 
and (\ref{23}) by the corresponding WKB-factors increase exponentially in
the semiclassical limit.

The $\chi$-factors of (\ref{22}) and (\ref{23}) satisfy the 
following 'initial' conditions:
\begin{eqnarray}
\chi_{1}^{\sigma}(x_0,x_0) = \chi_{2}^{\sigma{\prime}}(x_0,x_0) =1  &
\mbox{ and} &
\chi_{1}^{\sigma{\prime}}(x_0,x_0) = \chi_{2}^{\sigma}(x_0,x_0) = 0
\label{24}
\end{eqnarray}

\vspace{12pt}

\hskip+2em{\bf 2. Semiclassical expansions for $\chi_1 (x,\lambda)$ 
and $\chi_2 (x,\lambda)$  }
\vspace{12pt}

Consider now the solutions (\ref{22}) and (\ref{23}) defined at a vicinity of 
some point $x_0$.  We shall show below that if 
$x$ can be linked with $x_0$ by a canonical path the
solutions can be expanded semiclassically having the
corresponding forms (\ref{19}) where $x_0$ means now the 'initial' point
for the solutions. 

To formulate the corresponding lemma let us first invoke Theorem 4 of the
previous section to note that when $x_0$ is chosen then there are
always at least two fundamental solutions of opposite signatures
belonging to $N(x_0)$ which are Borel summable at the point $x_0$ and
communicate with themselves cannonically. Let us choose
these two fundamental solutions to be $\psi_a(x,\lambda)$ and
$\psi_b(x,\lambda)$.

 For the solution $\psi_1(x,\lambda)$ to the Schr\"odinger equation
(\ref{2}) defined by $\chi_1(x,\lambda)$ and the fundamental solutions
$\psi_a(x,\lambda)$ and $\psi_b(x,\lambda)$ we have: 
\begin{eqnarray}
\psi_1 (x,\lambda) =
\alpha(x_0,\lambda) \psi_a (x,\lambda) +
\beta (x_0,\lambda) \psi_b (x,\lambda)
\label{25}
\end{eqnarray}
due to the linear independence of the latter. For definitness 
we shall assume further that $\sigma_{1,2}=\sigma_a=-\sigma_b=-1$ in the
corresponding formulae for the solutions so that $\Re\xi(x)<\Re \xi (x_0)$ if
$x$ can be linked with $x_0$ by a canonical path. The coefficients $\alpha$
and $\beta$ in (\ref{25}) can be easily calculated according to the general
rules described in \cite{1,2}, for example. We have: 
\begin{eqnarray}
\alpha(x_0,\lambda)  =
\frac{\chi_b (x_0,\lambda)}{\chi_{a \rightarrow b}(\lambda)}
\exp \left (\lambda \int_{x_a}^{x_0}\sqrt{q(x,E)}dx \right) \nn \\
\beta (x_0,\lambda)  = \frac{1}{\chi_b (x_0,\lambda)}
\left( 1 - \frac{\chi_a(x_0,\lambda) \chi_b (x_0,\lambda)}
{\chi_{a \rightarrow b}(\lambda)} \right)
\exp \left (- \lambda \int_{x_b}^{x_0}\sqrt{q(x,E)}dx \right)
\label{26}
\end{eqnarray}
where the condition (\ref{24}) for $\chi_1(x,\lambda)$ has been used as well 
as the following relation 
\begin{eqnarray}
\chi_1 (\infty_b ,\lambda)  = \chi_b (x_0,\lambda)
\label{27}
\end{eqnarray}
The last relation generalizes a little bit a relation $\chi_{i\rightarrow j} =
\chi_{j \rightarrow i}$ valid for
any pair of fundamental solutions communicating canonically \cite{2}.  

We shall prove the following lemma.

\vskip 18pt
{\it\bf Lemma 3}

{\it  a) The factors $\chi_{1,2}$ given by}  (\ref{22}) $and$  (\ref{23}) 
{\it respectively can be expanded in corresponding domains $D_{1,2}=\{ x: 
\Re \xi(x)< \Re \xi(x_0)\}$
into the semiclassical series determined by the following
formulae }
\be
\chi^{as}_1(x,\lambda) = \frac{\chi^{as}_b(x_0,\lambda) 
\chi^{as}_a(x,\lambda)}{\chi_{a \rightarrow b}^{as}(\lambda)} = 
\frac{\chi^{as}_a(x_0,\lambda) \chi^{as}_b(x_0,\lambda)}
{\chi_{a \rightarrow b}^{as}(\lambda)}
\sum_{n\geq{0}} 
\left( \frac{\sigma}{2\lambda} \right)^{n}  I_n (x,x_0)
\label{28}
\ee

{\it and }
\be
\chi^{as}_2(x,\lambda) =\left(1 - \frac{\chi^{as}_a(x_0,\lambda) 
\chi^{as}_b(x_0,\lambda)}{\chi_{a \rightarrow b}^{as}(\lambda)}\right)
\frac{ \chi^{as}_a(x_0,\lambda)}{\chi^{'as}_a(x_0,\lambda)}
\sum_{n\geq{0}} 
\left( \frac{\sigma}{2\lambda} \right)^{n}  I_n (x,x_0)
\label{29}
\ee

{\it b) The domains $D_{1,2}$ are maximal for the above
expansions to be valid and are contained in the canonical domain
$D_a$ of the fundamental solution $\chi_a$.  

\vskip 6pt

c) The asymptotic series} (\ref{28}) $and$ (\ref{29}) {\it can be Borel 
summed with the following results }
\be
\left[ \chi_{1}^{as}(x,x_0,\lambda) \right]_{a}^{BS}  = 
C_{a}(x_0,\lambda)
\frac{\chi_{a}(x,\lambda)}{\chi_{a}(x_0,\lambda)} 
\label{30}
\ee

$ and$
\be
\left[ \chi_{2}^{as}(x,x_0,\lambda) \right]_{a}^{BS}  = 
(1 - C_{a}(x_0,\lambda))
\frac{\chi_{a}(x,\lambda)}{\chi'_{a}(x_0,\lambda)} 
\label{31}
\ee
{\it where the Borel sum $C_a(x_0,\lambda)\equiv
[\frac{\chi_a^{as}(x_0,\lambda)\chi_b^{as}(x_0,\lambda)}
{\chi_{a \rightarrow b}^{as}(\lambda)}]^{BS}$ is defined below.  
 
\vskip 6pt
d) The representations $(\ref{30})$ and $(\ref{31})$ are not unique.  }
\vskip 18pt 

{\it Proof of the lemma} 
\vskip 6pt

To prove the part a) of the lemma let us first 
devide both the sides of (\ref{25}) by  \\
$q^{-\frac{1}{4}}(x) \exp(-\lambda \int_{x_a}^{x}\sqrt{q(y)}dy)$ to get
\begin{eqnarray}
\chi_1 (x,\lambda) =
\frac{\chi_b(x_0,\lambda)}{\chi_{a \to b}(\lambda)} \chi_a (x,\lambda) +
\left(1-\frac{\chi_a (x_0,\lambda) \chi_b (x_0,\lambda)}
{\chi_{a \rightarrow b}(\lambda)} \right)
\exp\left(2\lambda \int_{x_0}^{x}\sqrt{q(y)}dy \right)
\frac{\chi_b(x,\lambda)}{\chi_{b}(x_0,\lambda)}
\label{32}
\end{eqnarray}

Next we note that the term in \mref{32}  proportional to $\chi_b(x,\lambda)$ 
 is exponentially small in
the semiclassical limit when compared with the first one. 
Therefore pushing $\lambda$ to infinity in (\ref{32}) we get 
(\ref{28}).

It is now easy to find the semiclassical series (\ref{29})
for $\chi_1(x,\lambda)$. To this end let us note that $\chi_{1,2}(x,x_0)$
are linear independent solutions of (\ref{15}) satisfying the
conditions (\ref{24}) so that we can write for $\chi_{a}(x,\lambda)$:
\begin{eqnarray}
\chi_{a}(x,\lambda) = 
\chi_{a}(x_0,\lambda) \chi_{1}(x,x_0,\lambda) +
\chi_{a}^{\prime}(x_0,\lambda) \chi_{2}(x,x_0,\lambda) 
\label{33}
\end{eqnarray}

Getting asymptotitcs of both the sides of (\ref{33}) and solving the 
obtained
equation with respect to $\chi_2^{as}(x,\lambda)$ we obtain (\ref{29}).

The thesis b) of the lemma follows
from the fact that both the solutions $\chi_{1,2}(x,\lambda)$ diverge
exponentially for $\Re x> \Re x_0$ when $\lambda \to \infty$ 
(the property which follows
directly when the considered pair of solutions is expressed by
the second pair of them defined by (\ref{22}) and (\ref{23}) with the
opposite signature) and from the fact that the condition $\Re x< \Re x_0$
defines also a (proper) part of the canonical domain $D_a$ of the
fundamental solution $\chi_a$.

  To prove the part c) of the lemma it
is necessary to invoke the exponential representation of the
fundamental solution $\chi$-factors \cite{2}. By this representation the
following is meant 
\begin{eqnarray}
\chi_{a,b}(x,\lambda) =
\exp\left(\mp \int_{\infty_{a,b}}^{x}\rho^{-}(y,\lambda)dy + 
\int_{\infty}^{x}\rho^{+}(y,\lambda)dy\right) \nonumber \\
\chi_{a,b}^{as}(x,\lambda) =
\exp\left(\mp \int_{\infty_{a,b}}^{x}\rho_{as}^{-}(y,\lambda)dy + 
\int_{\infty}^{x}\rho_{as}^{+}(y,\lambda)dy\right) 
\label{34}
\end{eqnarray}
\begin{eqnarray*}
\rho_{as}^{-}(y,\lambda) = \sum_{n \geq 0}\frac{\rho_{2n}^{-}(y)}
{{\lambda}^{2n+1}} \mbox{ ,  } \mbox{  }
\rho_{as}^{+}(y,\lambda) = \sum_{n \geq 0}\frac{\rho_{2n+1}^{+}(y)}
{{\lambda}^{2n+2}} 
\end{eqnarray*}
where the coefficients $\rho_{n}^{\pm}(y)$, $n \geq 0$, have been 
calculated explicitly in Ref. \cite{2}(see Sec.2.3 of this reference,
where the roles of $\rho^\pm(x,\lambda)$ are played by the
corresponding $\chi^\pm(x,E,\lambda)$-functions). 
The important properties of the coefficients as well 
as of the asymptotic series in (\ref{34}) they constitute are \cite{1,2}
 (see App.2 in Ref. \cite{2}): 

$a$. They are point (path independent) functions
of $y$, i.e. they are universal, sector independent functions; 

$b$. $\rho_{n}^{-}(y)$ have square root singularities at every turning point; 

$c$. $\rho_{n}^{+}(y)$ are meromorphic at each turning point with vanishing
residues at the points (i.e. $\oint \rho_{n}^{+}(y) dy =0$
around any turning point); 

$d$. Both the series in (\ref{34}) are Borel summable.  

This is the property $c$. which causes the $\rho_{as}^+$-integral in 
(\ref{34}) 
to be again the point function of $y$ i.e. it is sector independent.

The property $d$. which follows from the corresponding property of fundamental
solutions \cite{1} generates two Borel functions $\rho^{\pm}(x,s)$: 
\begin{equation}
\tilde{\rho}^{-}(x,s) = \sum_{n \geq 0}\frac{\rho_{n}^{-}(x)}
{(2n)!} s^{2n} \mbox{   ,  }
\tilde{\rho}^{+}(x,s) = - \sum_{n \geq 0}\frac{\rho_{n}^{+}(x)}
{(2n+1)!} s^{2n+1}
\label{35}
\end{equation}
which can be Borel 
transformed along any standard path $\tilde{C}$ in the Borel
plane providing us each time with the corresponding Borel summs
$\rho_{C}^{\pm}(x,\lambda)$ of the series in (\ref{35}). If we performed 
a Borel resummation of the 
first formula in (\ref{35}) along such a path $\tilde{C}$ we get: 
\begin{eqnarray}
\chi_{C}(x,\lambda) =
e^{\int_{\infty_{C}}^{x}(-\rho_{C}^{-}(y,\lambda) + 
\rho_{C}^{+}(y,\lambda))dy}
\label{36}
\end{eqnarray}
where $\chi_{C}(x,\lambda)$ is a $\chi$-function of some fundamental solution,
rotated possibly in the $\lambda$-plane. The minus sign in (\ref{36}) has
been chosen for definitness.

Noticing further, that: 
\begin{eqnarray}
\chi_{b}^{as}(x_0,\lambda)\chi_{a}^{as}(x_0,\lambda) =
e^{2\int_{\infty}^{x_0}\rho_{as}^{+}(x,\lambda)dx}\chi_{a\to b}^{as}(\lambda)
\label{37}
\end{eqnarray}
we can sum a la Borel both the equations (\ref{28}) and (\ref{29})
along the path $\tilde{C}_a$ recovering the factor $\chi_a(x,\lambda)$ 
to obtain the formulae (\ref{30}) and (\ref{31}). 
In these formulae $C_a(x_0,\lambda)$ is therefore
the following Borel sum 
\be
C_{a}(x_0,\lambda) =
\exp\left( 2\int_{\infty_a}^{x_0}\rho_{a}^{+}(y,\lambda)dy \right)
\label{38}
\ee

The representations (\ref{30}) and (\ref{31}) are
not unique since, in general, there can be other fundamental solutions in 
$N(x_0)$
with the same signature as the solutions $\chi_{1,2}$ have which can
substitute the solution $\chi_a$ in our considerations. However, 
having the same signatures these other fundamental solutions can provide us 
with the representations (\ref{30}) and (\ref{31}) which, in this case, 
differ between themselves only by exponentially small contributions.

The last statements ends our proof of Lemma 3. QED.

One can easily
identify the coefficients in front of the sum in the RHS's of
(\ref{28}) and (\ref{29}) as the corresponding series of constants in the
standard expansions (\ref{19}). It is important to note that $none$ of
them is equal to asymptotic series corresponding to 
$\chi_1(x_0,\lambda)\equiv 1$ and $\chi_2(x_0,\lambda)\equiv 0$ 
respectively. This confirms of course our earlier
statement that the semiclassical series (\ref{28}) and (\ref{29}) cannot be
Borel summed to the respective factors 
$\chi_1(x,\lambda)$ and $\chi_2(x,\lambda)$.
A reason for that is the presence of exponential terms
$e^{ -2\lambda \sigma(W_k (x) - W_k (x_0))}$ in the asymptotic 
formulae for (\ref{22})
and (\ref{23}) (when $\Re(W_k (x) - W_k (x_0))=0$) which breaks the necessary
conditions for the Watson-Sokal theorem \cite{5} to be applied. Note
that these exponential terms are absent in the case of fundamental
solutions which are obtained in the limit  $x_0 \rightarrow \infty_k$ taken
along a cannonical path, for any $k=1,2,...,2n+2$.

\section*{VI.  Uniqueness of fundamental solutions as Borel summable
solutions} 

\hskip+2em Let $\psi(x,\lambda)$ be any solution to the 
Schr\"odinger equation (\ref{2}) 
 given at some domain $D$ of the $x$-plane. Let us choose in $D$ a
point $x_0$ which is not a root of $q(x,E)$ (i.e. which is regular
for $\omega(x)$ as given by (\ref{6})). The solution 
$\psi(x,\lambda)$  can always be given each of
the two Dirac forms (\ref{3}) with the corresponding $\chi$-factors
satisfying the equation (\ref{15}). Let us write these forms in the
following way:
\be
\psi(x,\lambda) = C_{\pm}(\lambda)q^{-\frac{1}{4}}(x)
e^{\pm\lambda \int_{x_0}^x \sqrt{q(y)}dy}\chi_{\pm}(x,\lambda)
\label{39}
\ee

We shall say that $\psi(x,\lambda)$ has {\it well defined} semiclassical expansion
in $D$ if there is a choice of a sign in (\ref{39}) and an accompanied 
constant 
$C_{\pm}(\lambda)$ such that the $\chi$-factor $\chi_{\pm}(x,\lambda)$ 
corresponding to this choice can be expanded semiclassically in the 
standard way given by (\ref{19}).

It follows directly from the above definition that only one of the two 
possible choices can satisfy it.

We shall also say that $\psi(x,\lambda)$ is Borel summable in $D$ if it has 
well defined 
semiclassical expansion there and the corresponding series (\ref{19}) is 
Borel summable to the uniquely chosen $\chi$-factor $\chi (x,\lambda)$ of 
$\psi(x,\lambda)$.

We shall prove below the following main theorem of this paper:

\begin{moje}  
Let a solution $\psi(x,\lambda)$ given at some vicinity $D$ of $x_0$
($x_0$ does not coincide with any turning point) be Borel summable in 
$D$. Then this solution must coincide with one of the
fundamental solutions up to some $\lambda$-dependent constant .
\end{moje}

$Proof.$

To prove the theorem we could utilize the solutions (\ref{22})
and (\ref{23}) and all their properties which we have established in
Lemma 3 of the previous section. It can however be quite
instructive to prove the theorem not invoking for the latter
solutions since it makes the main arguments supporting the
theorem (which have worked implicitly also in proving Lemma 3 of
the previous section) to be more transparent.

According to Theorem 4 of Section III., for $x_0$ chosen
we can always find in the set $N(x_0)$ a number of fundamental
solutions of the same signatures as the respective $\chi(x,\lambda)$ 
corresponding to $\psi(x,\lambda)$ has. Let 
$\chi_a(x,\lambda)$ be one of them. 
It is Borel summable at $x_0$ and in some of its 
vicinity. Then using (\ref{20}) both for $\chi(x,\lambda)$ and
$\chi_a(x,\lambda)$ we have: 
\begin{eqnarray}
\frac{\chi^{as} (x,\lambda)}{\chi^{as} (x_0,\lambda)} =
\sum_{n\geq{0}} 
\left( \frac{\sigma}{2\lambda} \right)^{n}  I_n(x,x_0) =
\frac{ \chi_a^{as} (x,\lambda)}{ \chi_a^{as} (x_0,\lambda)}
\label{40}
\end{eqnarray}

It follows from (\ref{40}) that the outer parts of
this equality having the same semiclassical expansions have to
have also the same Borel function. Since $\chi^{as}(x,\lambda)$ and 
$\chi^{as}(x_0,\lambda)$ are both Borel summable in $D$ they can be summed 
along the same standard path $\tilde{C}$ on their corresponding Borel 
planes if $x$ 
is chosen to be sufficiently close to $x_0$. It is, however, easy to check
(see App.  3) that under the latter condition the same standard path 
$\tilde{C}$ can be chosen to sum the quotient on the LHS of (\ref{40}) since
its corresponding Borel function is holomorphic around this
path.  However, the same must be true for the RHS quotient i.e.
the corresponding Borel functions of its two factors can be
integrated also along $\tilde{C}$ lying in their Borel planes. Let us sum
therefore a la Borel both the outer sides of (\ref{40}) along this
path. We get 
\begin{eqnarray}
\chi (x,\lambda)  = \chi (x_0,\lambda) 
\frac{2\lambda \int_{\tilde{C}} e^{2 \lambda s} \tilde{\chi}_a(x,s) ds}
{2\lambda \int_{\tilde{C}} e^{2 \lambda s} \tilde{\chi}_a(x_0,s) ds}
\label{41}
\end{eqnarray}

The last equation, however, ends the proof of the
theorem. QED.  

As a comment to the last theorem we would like to
stress that it summarizes a particular property of the
semiclassical theory of the 1D Schr\"odinger equation with the
polynomial potentials. Namely, this is that the standard
semiclassical expansion (\ref{19}) is constructed basically by the series 
$\sum_{n\geq{0}} \left( \frac{\sigma}{2\lambda} \right)^{n}  I_n(x,x_0)$
which can be Borel summable and the Borel function of
which, by (\ref{40}), coincides up to a $\lambda$-dependent multiplicative
constant with the one of the fundamental solutions and, also by
(\ref{40}), with the Borel function of any Borel summable solution.
This means that we can consider the Borel function of the
fundamental solutions as the canonical one. The latter can be
uniquely defined by the condition of being equal to unity at $s=0$
on the 'first sheet' of the corresponding Riemann surface which
the condition it satisfies actually.

\section*{VII.  Conclusions and discussion} 

\hskip+2em Theorem 5 of the previous section
shows that in the case of the Schr\"odinger equation with the
polynomial potentials its Borel summable solutions are the
fundamental ones. The Borel function generated by these
solutions is, up to analytical continuation, the unique one.
This property justifies our earlier use of the fundamental
solutions to investigate the problem of the Borel summability of
energy levels and matrix elements in 1D quantum mechanics \cite{1}.
It shows also that only the fundamental solutions can be invoked
when $any$ problem connected with the Borel resummation is
considered and conditions for such resummations are satisfied \cite{5}.  

The latter objection is important since not all the
results we obtain for the case of polynomial potentials can be
immediately extended to other cases of potentials. These are,
for example, the rational potentials being the next class of
potentials of the modeling importance.  In particular, the
universality of the Borel function in the later case of
potentials seems to be not satisfied \cite{8}.  

Nevertheless, the
role of the corresponding fundamental solutions as the unique
Borel summable ones seems to be maintained not only in the case
of rational potentials but also in the case of other meromorphic
potentials such as the P\"oschl-Teller one, for example.  

The fundamental solutions we have discussed in Sec. 2 can be given
another forms when each of the factors in (\ref{3}) becomes a
complicated function of $\lambda$ \cite{9}. These generalized 
representations
however preserve all the Borel summing features of the original
fundamental solutions being only a partial Borel resummation of
the latter \cite{10}.

Finally, we would like to note that the result obtained in the present 
paper completes the ones obtained in our other papers \cite{1,6,10}. 
Namely, all these results show that the semiclassical theory in 1D
 quantum mechanics can be completely formulated on the base of the Borel
method of resummation. This is certainly true in the case of the 
polynomial potentials and it seems to be true with some modifications 
for meromorphic potentials as well \cite{8}. In the formulation of such 
a theory the essential role as we have shown in the present paper is 
played by the fundamental solutions (see also \cite{1,6,10}). The theory 
allows us to construct the simplest semiclassical approximations as well 
as to complete the latter by the exponentially small contributions up to
a desired level of accuracy \cite{6}. In such a theory even a change of 
variable in the Schr\"odinger equation the procedure which is used very 
frequently as a way of improving the semiclassical approximations is 
also a result of the proper Borel resummation operation \cite{10,9}.

\vskip 12pt

{\large {\bf Acknowledgments}}

Stefan Giller has been supported by the KBN grant
2PO3B 07610 and Piotr Milczarski  by
the \L\'od\'z University Grant No 795.

\vskip 3pt
\
\section*{Appendix 1}
\hskip+2em Here we would like to draw some basic conclusions which follow 
for the Borel function $\tilde\chi(x,s)$
 from its representation given by the topological expansion developed in our 
recent paper
(see \cite{6}) and not dicussed there.

First of all let us recapitulate shortly basic elements of this 
representation. Namely, we
have shown in \cite{6} that the Borel function  defined originally in some 
sector, say $S_1$,
can be represented in this sector as the following series 
($\xi=\xi(x)=\int_{x_1}^x\sqrt{q(y,E)}dy$):
\be
\tilde{\Phi}_1(\xi,s)&= \sum_{q\geq 0} \tilde{\Phi}_1^{(q)}(\xi,s)\\
\label{A1.11}\nn
\ee
where $\tilde{\Phi}_1(\xi(x),s)\equiv \tilde{\chi}(x,s)$ and the terms 
$\tilde{\Phi}^{(q)}(\xi,s), q \geq 0$  of the series are given by the formulae:
\begin{eqnarray}
\label{A1.12}\nn
\tilde{\Phi}_1^{(0)}(\xi,s)&=& I_0 \left(\sqrt{4s\Omega(\xi)}\right)\nn\\ 
\tilde{\Phi}_1^{(2q)}(\xi,s)&=& \int\limits_{\tilde{C}(s)} d\eta_1 
\int\limits_{\tilde{C}(\eta_1)} d\eta_2  \ldots
\int\limits_{\tilde{C}(\eta_{q-1})} d\eta_q 
\int\limits_{\infty_1}^{\xi-\eta_1} d\xi_1
\int\limits_{\infty_1}^{\xi_1} d\xi_2 \ldots
\int\limits_{\infty_1}^{\xi_{q-1}} d\xi_q \nn \\
&&\tilde{\omega}(\xi_1 +\eta_1)\tilde{\omega}(\xi_1 +\eta_2) \cdot \ldots \cdot\tilde{\omega}(\xi_q +\eta_q)\tilde{\omega}(\xi_q)(2s-2\eta_1)^{2q}
\frac{I_{2q}(z_{2q}^{\fr})}{z_{2q}^q} \nn \\
& & \nn\\
z_{2q}&=&4(s-\eta_1)\Omega(\xi) + 8(s-\eta_1) \sum_{p=1}^q
\left( \Omega(\xi_p + \eta_{p+1}) - \Omega(\xi_q + \eta_p) \right), 
 \nn \\
\eta_{q+1} &\equiv & 0, \;\;\;\;\;\;\;\;\;\; q=1,2,\ldots \\
& &\nn\\
\tilde{\Phi}_1^{(2q+1)}(\xi,s)&=& 
 \int\limits_{\tilde{C}(s)} d\eta_1  \ldots
 \int\limits_{\tilde{C}(\eta_{q})} d\eta_{q+1} 
\tilde{\omega}(\xi-\eta_1+\eta_2)
 \int\limits_{\infty_1}^{\xi-\eta_1} d\xi_1 \ldots
\int\limits_{\infty_1}^{\xi_{q-1}} d\xi_q  \nn \\
&&\tilde{\omega}(\xi_1 +\eta_2)
\tilde{\omega}(\xi_1 +\eta_3) \cdot \ldots \cdot
\tilde{\omega}(\xi_q +\eta_{q+1})\tilde{\omega}(\xi_q)(2s-2\eta_1)^{2q+1}
\frac{I_{2q+1}(z_{2q+1}^{\fr})}{z_{2q+1}^{\frac{2q+1}{2}}}  \nn \\
z_{2q+1}&=& 
4(s-\eta_1)\Omega(\xi) + 8(s-\eta_1) \sum_{p=0}^q
\left( \Omega(\xi_p + \eta_{p+2}) - \Omega(\xi_p + \eta_{p+1}) \right), 
 \nn \\
\xi_0 &\equiv & \xi ,\;\;\;\;\;\eta_{q+2} \equiv 0 , \;\;\;\;\; q=0,1,2,...\;\;\;\;\;\;\;\;\;\; \nn 
\end{eqnarray}
where $\tilde{\omega}(\xi(x))\equiv\omega(x)q^{-\fr}(x), 
\Omega=\int\limits_{\infty_1}^{\xi}\tilde{\omega}(\eta)d\eta$ and the 
functions $I_q(x), q \geq0$, in (\ref{A1.12}) are the
modified Bessel functions (of the first kind, \cite{Bate} p.5, formula (12)).

The formulae (\ref{A1.12}) can be obtained from the following recurrences:
\be\label{A1.14}
\tilde{\Phi}_1^{(2q+2)}(\xi,s)&=&- \int\limits_{\tilde{C}(s)} d\eta \int\limits_{\tilde{C}(\eta)} d\eta' \int\limits_{\infty_1}^{\xi}d\xi_1\tilde{\omega}(\xi_1)\tilde{\omega}(\xi_1 -\eta')(2s-2\eta)\nn\\ 
&&\tilde{\Phi}_1^{(2q)}(\xi_1 -\eta',\eta- \eta')\frac{I_{1}\left(
\sqrt{4(s-\eta)(\Omega(\xi)-2\Omega(\xi_1)+\Omega(\xi_1-\eta'))}\right)}
{\sqrt{4(s-\eta)(\Omega(\xi)-2\Omega(\xi_1)+\Omega(\xi_1-\eta'))}} \nn \\
&&\\
\tilde{\Phi}_1^{(2q+1)}(\xi,s)&=&- \int\limits_{\tilde{C}(s)} d\eta
 \int\limits_{\tilde{C}(\eta)} d\eta'
\tilde{\omega}(\xi-\eta')\nn\\
&&\tilde{\Phi}_1^{(2q)}(\xi -\eta',\eta-\eta')
I_{0}\left(\sqrt{-4(s-\eta)(\Omega(\xi)-\Omega(\xi-\eta'))}\right)\nn\\ 
&&\nn\\
\;\;\;\;\;\;\;\;\;\;\;\;\;\;\;  q&=&0,1,2,... \nn
\ee
where $\tilde{\Phi}_1^{(0)}(\xi,s)$ is given by (\ref{A1.12}).

Note that (\ref{A1.14}) can be obtained from (\ref{A1.12}) and 
vice versa by applying the following relations:
\be\label{A1.15}
 \int\limits_0^1 dx I_m(\sqrt{\alpha x})I_m(\sqrt{\beta(1- x)})
(\alpha x)^{\fr m}(\beta(1-x))^{\fr n}&=&
2\alpha^m \beta^n \frac{I_{m+n+1}(\sqrt{\alpha+\beta})}
{(\sqrt{\alpha+\beta})^{m+n+1}} \nn\\
&&\\
\frac{1}{(k-1)!(n-k)!}
\int\limits_{\eta}^s d\eta'(s-\eta')^{k-1}(\eta'-\eta)^{n-k}&=&\frac{(s-\eta)^n}{n!} \nn
\ee

The $\xi$-integrations in (\ref{A1.12}) and (\ref{A1.14}) run over some 
$\xi$-Riemann surfaces of the
subintegral functions starting from the infinite points of these surfaces 
which are the
corresponding images of the infinite point of the sector $S_1$. The 
$\eta$-integrations, contrary to the
$\xi$-ones, are finite and run over the s-Riemann surfaces. All the latter 
integrations ends at $s=0$.

For these integrations the most important are the branch point structures of 
the Riemann
surfaces corresponding to the functions $\tilde{\omega}(\xi)$ and 
$\Omega(\xi)$ and the shifts of these surfaces by some
complex number $\eta$. The two latter surfaces corresponds to the functions 
 $\tilde{\omega}(\xi-\eta)$ and $\Omega(\xi-\eta)$.

Since every of these four surfaces has complicated topology (defined by its 
branch points)
we decided not trying to sew them suitably together when these functions are 
integrated simultaneously
but rather to consider them, for safeness, separately. These topologies are 
determined, of course,
by the singularities of the respective functions $\tilde{\omega}(\xi)$ and 
$\Omega(\xi)$. Besides, since the latter of these
two functions is defined as the integral over the former then the 
corresponding Riemann surface
on which $\Omega(\xi)$ is defined is a map of the surface corresponding to 
$\tilde{\omega}(\xi)$, i.e. there 
is a well
defined relation between these two surfaces.

For the polynomial potentials with simple roots all singularities of $\tilde{\omega}(\xi)$ and 
$\Omega(\xi)$ are cubic root branch
points corresponding to turning points. Therefore, the four surfaces 
discussed above acquires a
suitable cut pattern each. The corresponding $\xi$-integration runs over the 
sheets of these surfaces
which are unambiguously related to each others (by the above shift or by a 
map) so that the
integration paths on this sheets look the $same$ runnig from $\infty_1$ 
to some finite point, the $same$ on each sheet.

It is necessary to stress that every subsequent integration in (\ref{A1.12}) 
or every subsequent
step in the recurrent formulae (\ref{A1.14}) changes the structure of the 
Riemann surfaces
corresponding to functions resulting from these integrations. Namely, these 
surfaces become still
more complicated preserving all the branch points of the previous stage and 
acquiring new ones
as a result of the last integration(s).

Nevertheless, these structures look relatively simple if we consider them on 
definite
sheets of the Riemann surfaces we want to stay considering the properties of 
the Borel functions $\tilde{\Phi}(\xi,s)$. Namely, starting from the sheets 
corresponding to the sector $S_1$ we shall keep the
variable $x$ (or $\xi(x)$) changing along the contour $K'$ of Fig.1a. 
This corresponds to a path $\gamma_1(x)$ on the $x$-plane which
begins in the sector $S_1$ and crosses on its way all the Stokes lines 
running to the infinity of the
plane (but each line only once) penetrating subsequent sectors in their 
cyclic (clockwise or anticlockwise) ordering introduced in Section II. 
We shall call such a path the outer path.

Under the above condition the final pattern of the branch points of 
$\tilde{\Phi}(\xi,s)$ viewed from
the relevant sheets is quite simple. First let us note that, as it follows 
from (\ref{A1.12}) and (\ref{A1.14}),
under the above circumstances we can deform homotopically the infinite end 
of the outer path $\gamma$
from the sector $S_1$ to any of the subsequent sectors $S_2,...,S_{n+1}$ 
(in this or in the reversed orders).
This is the consequence of the fact that the initial condition (\ref{7}) put 
on the path $\gamma$ to make it
canonical is no longer valid since all the dangerous exponentials in the 
formula (\ref{5}) enforcing this
condition disapeared on the way of passing to the formulae (\ref{A1.12}) and 
(\ref{A1.14}). Fig.6 shows the corresponding result of such an
operation for $x\in K'\cap S_3$. This proves
that under the above condition (\ref{A1.12}) and (\ref{A1.14}) define the 
$same$ unique Borel function $\tilde{\Phi}(\xi,s)$
 for $all$ the $n+1$ fundamental solutions (\ref{3}). This deformation can be
 done keeping the end
point $x$ ($\xi(x)$) of the outer path $\gamma$ in $any$ of the sectors 
$S_1,...,S_{n+1}$. The latter property means,
of course, that if $\xi$ is in $S_k$ and the infinite end of $\gamma$ is in 
$S_l$ then $\tilde{\Phi}(\xi,s)$ represents the Borel
function of the fundamental solution $\psi_k$ (defined in $S_k$) analitically
 continued to the sector $S_l$ and
the formulae (\ref{A1.12}) and (\ref{A1.14}) define then this continuation 
explicitly.

\vskip 15pt

\begin{tabular}{c}
\psfig{figure=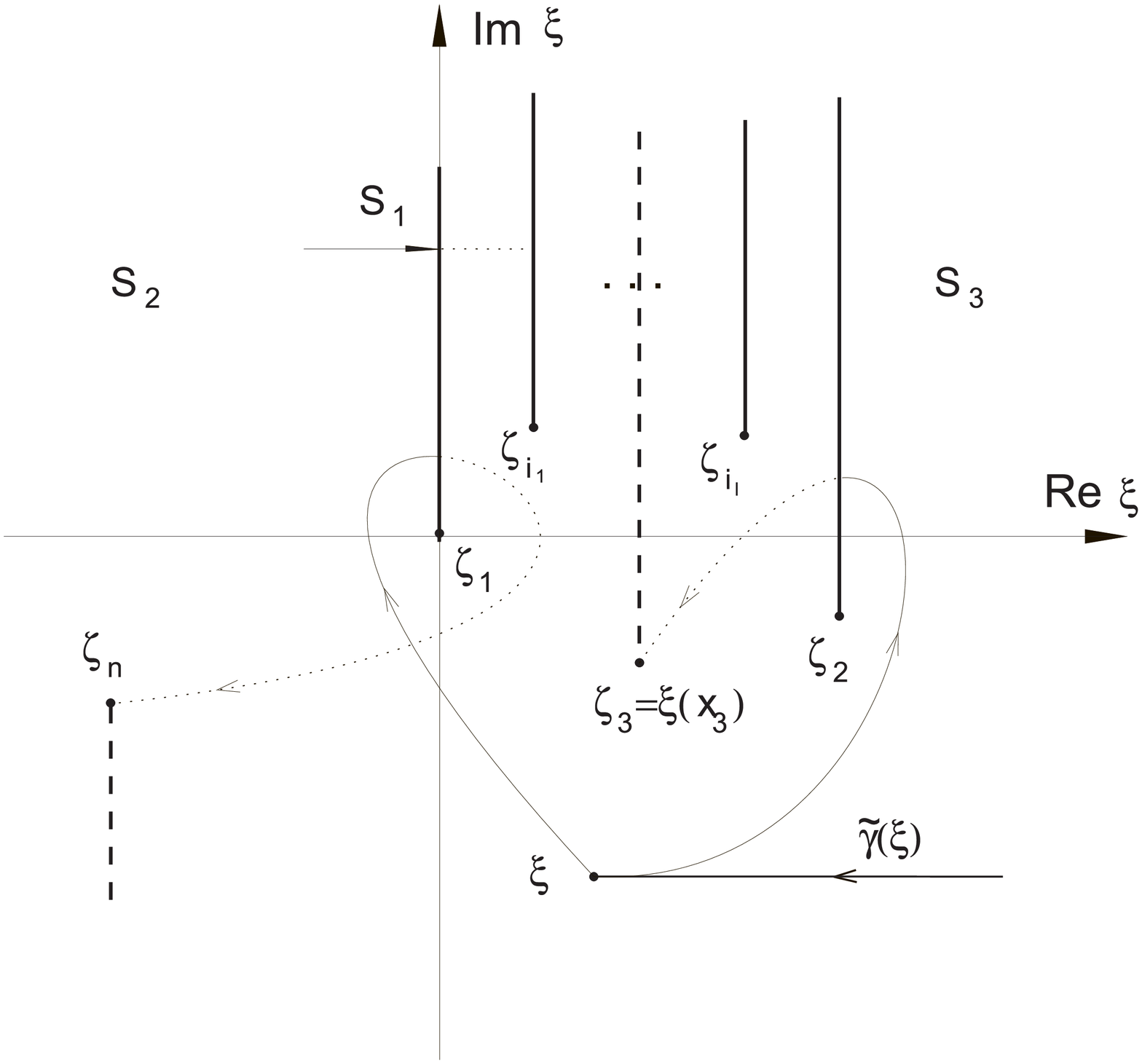,width=10cm} \\
Fig.6 The branch point structure of the $\xi$-Riemann surface
  sheet containing \\the sectors $S_2$ and $S_3$ for
  $\tilde\Phi(\xi,s)$ when $x\in K'\cap S_3$ 
\end{tabular}

\vskip 15pt

The following theorem can now be proved inductively.

\begin{moje}
Let the end point $x$ go around the contour $K'$ of $Fig.1a$ starting from the
sector $S_1$ and passing consecutively by the sectors $S_2,...,S_{n+1}$. Then
 the sheets from which
the $\chi$-factors corresponding to the subsequent sectors are recovered by 
the Borel
transformations over the Borel function $\tilde{\chi}(x,s)$ along the 
suitable real halfaxes have the
branch point structures shown in $Fig.11$. 
\end{moje}

\vskip 6pt     

$Proof$.

Let us consider first a sheet of the s-Riemann surface corresponding to the 
sector
$S_1$ and let the point $x$ be then on $K'$ in the sector $S_1$ so that 
$\Re \xi(x)>0$.

Consider $\tilde{\Phi}^{(0)}(\xi,s)$. Since $I_0(\sqrt{4s\Omega(\xi)})$ is 
the holomorphic function of its argument then $\tilde{\Phi}^{(0)}(\xi,s)$
is an entire function of $s$ (i.e. holomorphic in the whole $s$-plane) so 
that its corresponding $s$-Riemann surface coincides with the $s$-plane. What concerns its $\xi$-Riemann 
surface structure it
coincides with the one of $\Omega(\xi)$ since for each natural power of the 
latter its $\xi$-Riemann structure
is the same and $I_0(\sqrt{4s\Omega(\xi)})$ determines $\tilde{\Phi}^{(0)}(\xi,s)$
 as the holomorphic function of $\xi$ in each non-
singular point of $\Omega(\xi)$. Therefore $\tilde{\Phi}^{(0)}(\xi,s)$ has on 
the corresponding first 
sheet in the $\xi$-plane
a unique branch point at $\zeta_1$, as it is shown in Fig.7a, if the 
corresponding cut emerges
from this branch point vertically down. There are no other branch points 
visible then on the
sheet. These another branch points, however, are on the sheets lying $below$ 
the first sheet and
the closest ones at $\zeta_2, \zeta_{i_1},..., \zeta_{i_l}$ and $\zeta_n$ 
are shown in Fig.7a with the dashed lines of cuts emerging
from them. The full thin paths on the figure emerging from the point $\xi$ 
show how to approach these last
branch points starting from $\xi$. We shall adopt this convention for the 
remaining figures too.

Fig.7b shows the $\xi$-Riemann surface for the shifted function 
$\tilde{\Phi}^{(0)}(\xi-\eta,s)$.

\vskip 15pt

\begin{tabular}{cc}
\psfig{figure=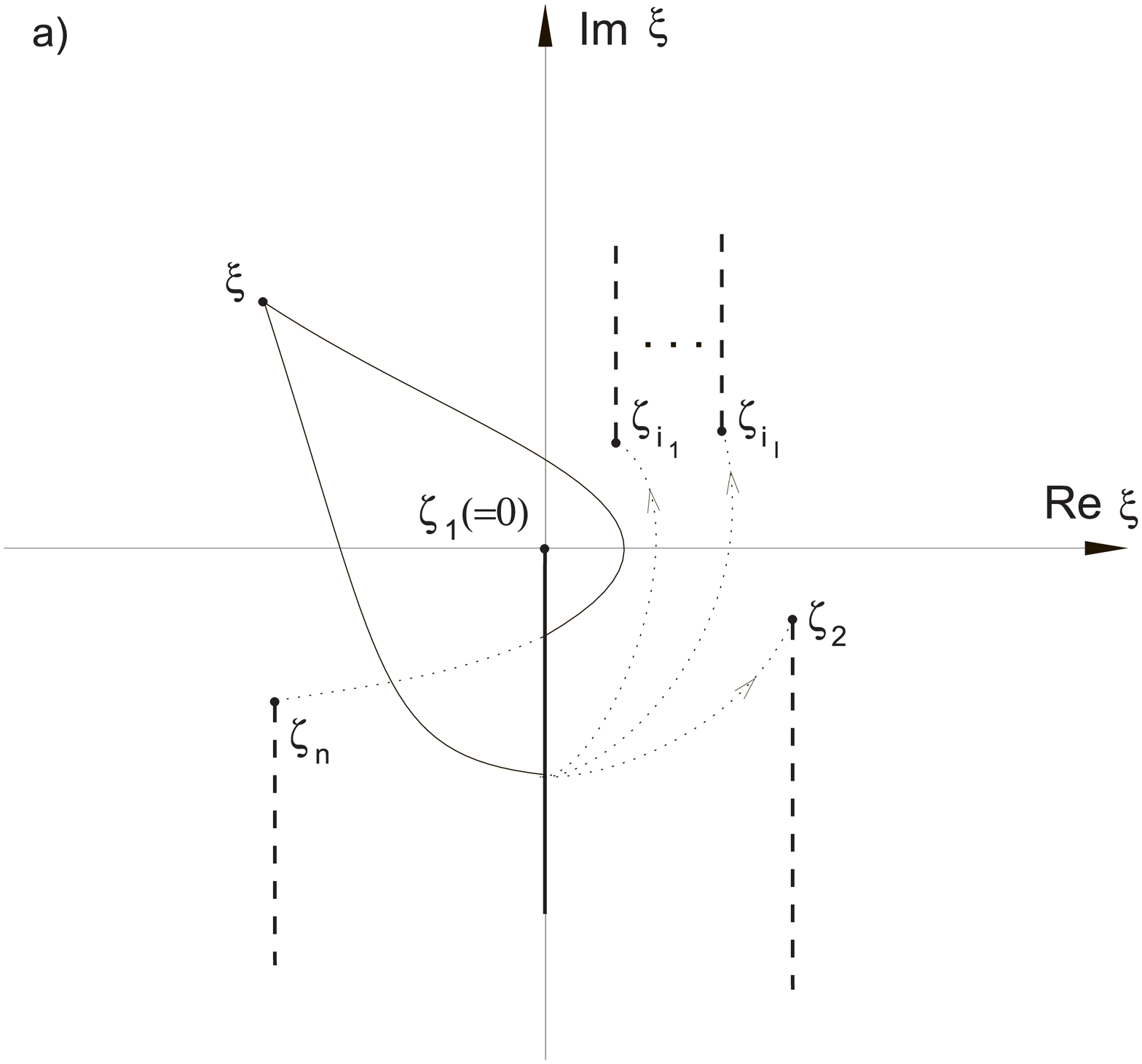,width=6.5cm} & \psfig{figure=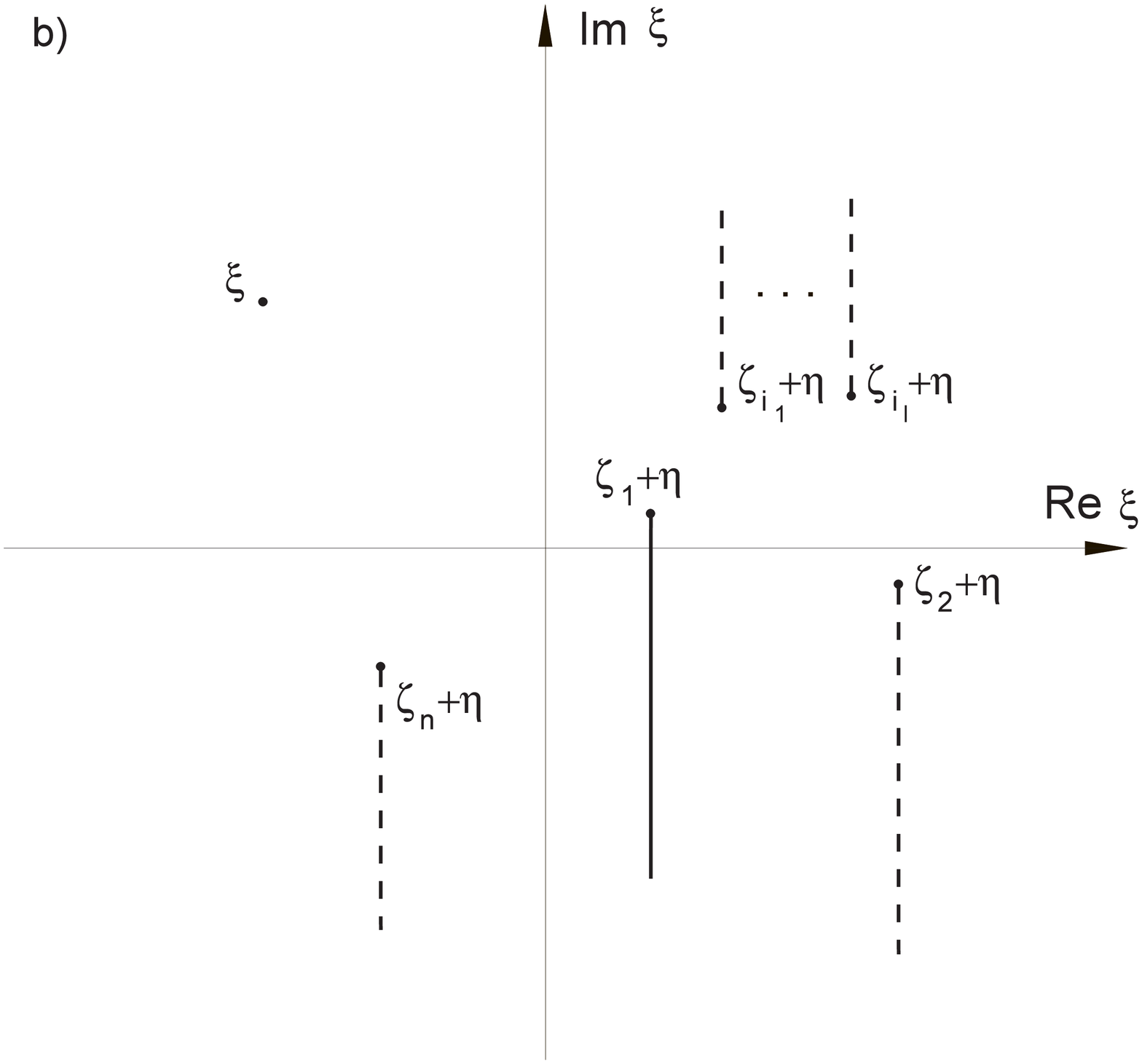,width=6.5cm} \\
Fig.7a The 'first' sheet of the $\xi$-Riemann&  Fig.7b The 'first'
sheet of the $\xi$-Riemann\\surface corresponding to
$\tilde\Phi^{(0)}(\xi,s)$&surface corresponding to $\tilde\Phi^{(0)}(\xi-\eta,s)$\\when $x\in K'\cap
  S_3$&when $x\in K'\cap S_3$
\end{tabular}

\vskip 15pt

Next consider $\tilde{\Phi}^{(1)}(\xi,s)$. It is given by the second of the 
formulae (\ref{A1.14}) for $q=0$. In
this formula we have to integrate first over the variable $\eta'$. As it 
follows from the formula the
branch point structure of the $\eta'$-Riemann surface is determined by the 
functions $\tilde\omega(\xi-\eta')$, $\Omega(\xi-\eta')$ 
and $\tilde{\Phi}^{(0)}(\xi-\eta',\eta'-\eta)$ and its first and the second 
sheets are shown on Fig.8a.

The integration over $\eta'$ leads us to a function defined on the 
$\xi$-Riemann function shown
in Fig.7b. The cuts shown there are a result of the end point (EP) 
mechanism of the
singularity producing (\cite{6}, see also \cite{Polking}). Namely, the $\eta'$-integration is perturbed if the 
moving branch points of
Fig.8a can approach the fixed end points $\eta'=0$ and $\eta'=\eta$ 
of the integration path $C(\eta)$. For
example, to generate the branch point at $\zeta_1(=0)$ we simply move the 
branch point $\xi-\zeta_1$ against
the end point $\eta'=0$ of $C(\eta)$ to touch it finally. To produce the 
branch point at $\zeta_2$ on Fig.8b
we have to move $\xi-\zeta_1$ down avoiding the end point $\eta'$ from the 
left and below and next
moving it to the right in such a way to make the screened branch point
at $\xi-\zeta_2$ coinciding with the end point $\eta'$.

\vskip 15pt

\begin{tabular}{cc}
\psfig{figure=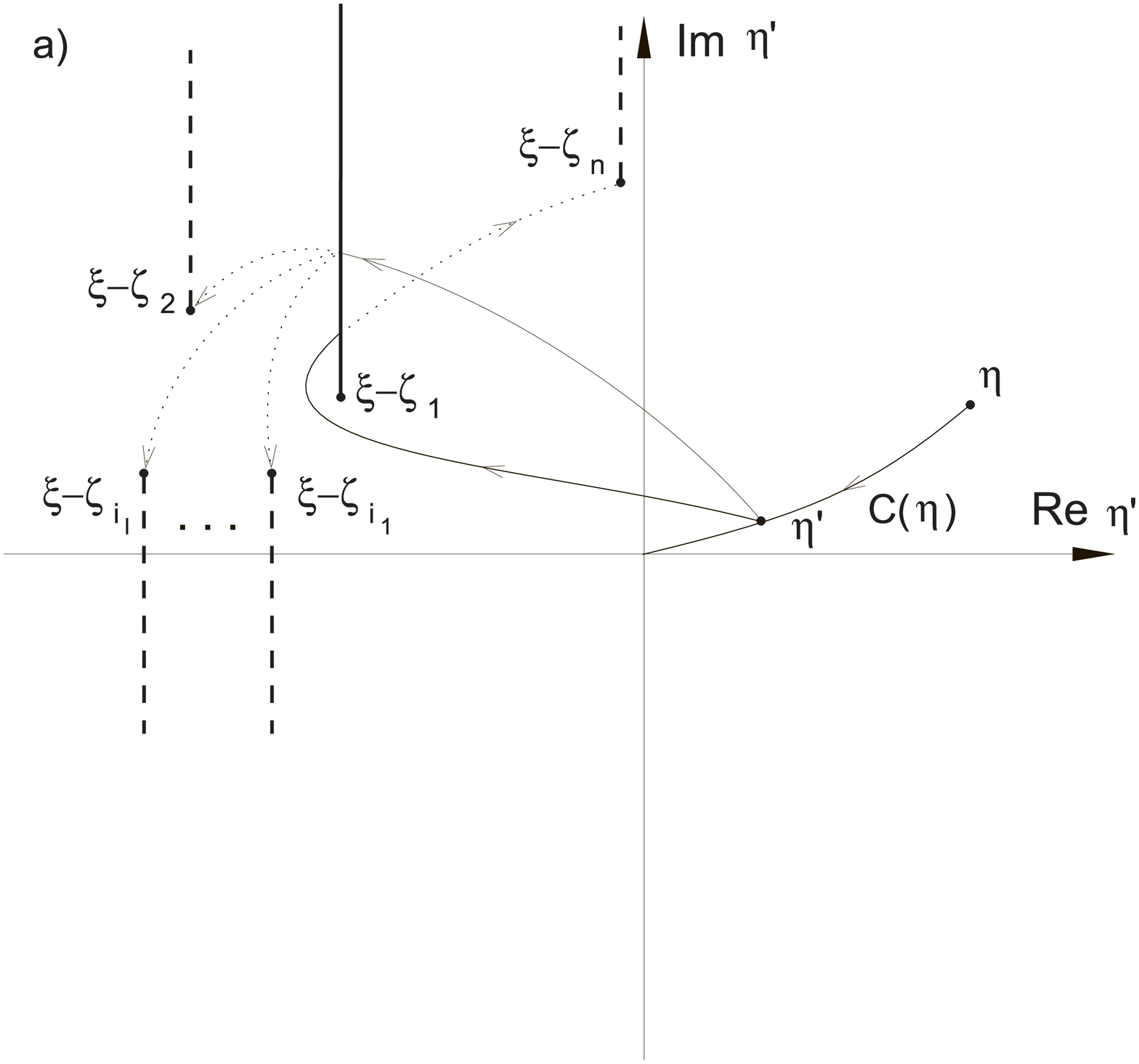,width=6.5cm} & \psfig{figure=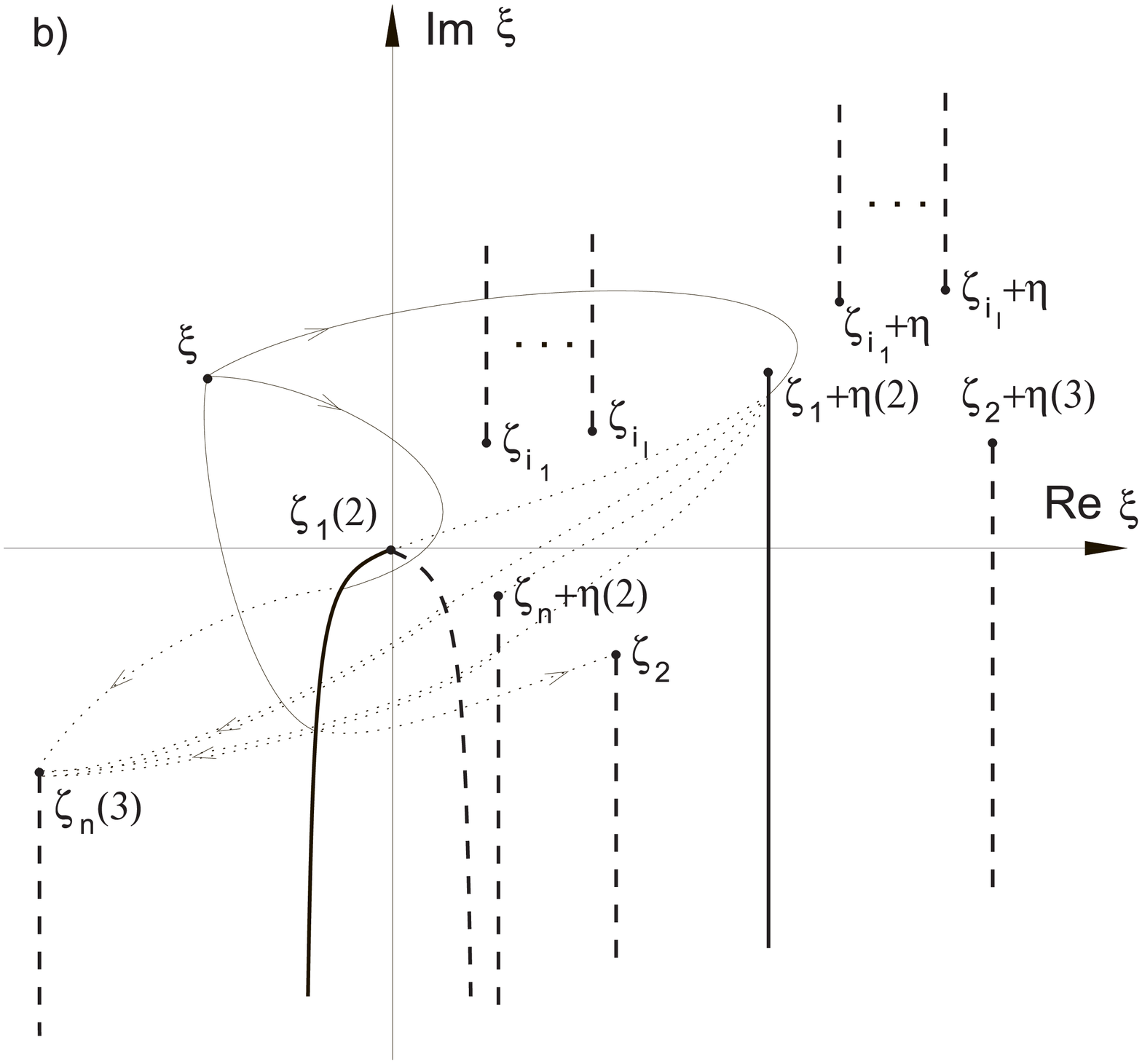,width=6.5cm} \\
Fig.8a The singularity structure of the&Fig.8b The resulting $\xi$-singularity structure\\integrand in the
  second formula (48) for&of $\tilde\Phi^{(1)}(\xi,s)$\\$q=0$ and the integration path $C(\eta)$&$\;$
\end{tabular}

\vskip 15pt

In the way described above we can generate the branch points at the position 
shown in
Fig.8b. A convention adopted on this figure for the cut designing is 
to draw a full thin path emerging from the point $\xi$ and if the path 
crosses a cut it becomes dotted. If it crosses the next appropriate cut it is
doubly dotted and so on. This way of designing cuts allows us to establish 
the sheets they are distributed on. The figure in parentheses at the 
different branch points indicate the multiplicity of the latter i.e. their 
appearing on $different$ sheets at the $same$ positions. However, for the
sake of transparency of the figures not all paths showing the distributions of 
the branch points on the sheets have been shown.

It is worth to note that this multiplication of branch points 
with the same coordinates
but lying on different sheets on this figure is necessary to keep fixed the 
relative distribution of
the branch points on Fig.8a when the moving ones on Fig.8b 
change their positions. In fact
the branch points in this figure with the same coordinate substitute each 
other during such a
motion. This is because the branch point $\xi-\zeta_1$ on Fig.8a 
has to be always accompanied by the
branch points at $\xi-\zeta_2$ and $\xi-\zeta_n$ lying on the next two lower 
sheets. For example, if the branch
points $\zeta_1+\eta$ and $\zeta_2+\eta$ move to the left so that they pass 
the branch point at $\zeta_1(=0)$ having it
between themselves then the point $\zeta_2+\eta$ on the sheet opened by the 
first of these moving points
is screened in some moment by the cut emerging from $\zeta_1=0$ lying on the 
first sheet in Fig.8b.
But then the branch point at $\zeta_2+\eta$ on the sheet opened just by this 
latter cut becomes unscreened
substituting its copy on the sheet we started with. Of course, all three 
copies of the branch points
at $\zeta_2+\eta$ are mapped by the relation $\xi-\eta=\zeta_k$, $k=1,2,3$, 
into the one copy of them at $\xi-\zeta_2$ on the $\eta'$-plane of 
Fig.8a.

The final integration over $\eta$ repeats only the steps done earlier 
introducing nothing new
to the distribution of the branch ponts on the first sheet not modifying the 
lower sheets as well
so that the final branch point structures of the first sheets look again 
as in the figures 8a-b where we have to substitute $\eta'$ and $\eta$ 
by $s$ on both the figures.

Consider next $\tilde{\Phi}^{(2)}(\xi,s)$. It is given by the first of the 
formulae (\ref{A1.14}) for $q=0$. The
initial $\xi$-Riemann surface structure is similar to that shown on 
Fig.7 and look as in Fig.9a-b.
The first $\xi_1$-integration in the corresponding formula (\ref{A1.14}) 
provides us this time with both
the types of singularities on the $\eta'$-Riemann surface i.e. generated by 
the E-mechanism, which
makes a replica of the branch point structures of Fig.8a, and by 
the pinch (P) mechanisms
(discussed in \cite{6}, see also \cite{Polking}). A singularity generated by 
the second mechanism arises when, for
example, the branch point at $\zeta_1+\eta'$ on Fig.9b moves against 
the one at $\zeta_1=0$ of Fig.9a
pinching the integration path $\tilde{\gamma}(\xi)$. It can be done by making 
a tour around the end $\xi$ of the path $\tilde{\gamma}(\xi)$
in both the directions i.e. clockwise anticlockwise. In this example we will 
produce branch point
singularities on both the lower sheets of the $\eta'$-Riemann surface shown 
if Fig.9c at $\eta'=0$. We
can pinch the path $\tilde{\gamma}(\xi)$ by $\xi_1+\eta'$ also in this way 
but against the branch points at $\zeta_2$ and $\zeta_n$.
This needs only to cross the cut emerging from $\zeta_1(=0)$ to reach the 
points mentioned i.e. the first one by crossing this cut from the left whilst 
the second - from the right. It means that $\eta'$
itself has to do the same on Fig.9c crossing the cuts emerging from 
$\eta'=0$ on the corresponding
sheets. The positions of the branch points generated in this way are shown 
on Fig.9c where only a closest part of them is shown. It
follows from the figure that some of the closest fixed branch points are 
at $\eta'$ on the two sheets opened
by the branch point at $\xi-\zeta_1$. The other ones lie on the sheets 
opened by the the branch
points at $\xi-\zeta_2,\xi- \zeta_{i_1},...,\xi- \zeta_{i_l}$ and $\xi-\zeta_n$ and 
by the fixed branch points 
generated in this way i.e. still on the lower sheets.

A discussion of the remaining two $\eta'$- and $\eta$-inegrations goes along 
the same lines as in
the previous discussion on calculating $\tilde{\Phi}^{(1)}(\xi,s)$ with the 
similar results obtained accordingly to
Fig.9c. The only appearing difference is 
that in both 
these integrations the P-mechanism of the moving branch point singularity 
generation on the $\xi$-
Riemann surface becomes active since, except the moving singularities, there 
are also the fixed
ones on the corresponding sheets of the $\eta'$- and $\eta$-Riemann surfaces 
(see Fig.9c). Therefore, the final first sheet structures are again a
replica of those shown in Fig.8b for the $\xi$-Riemann surface and in
Fig.9c for the $s$-Riemann one (with the $\eta,\eta'$-variables
substituted suitably by the $s$-one). Of course,
because of the reason of transparency only the branch points on the first 
three sheets are shown
on the figures. A comment made previously on the proliferation of the moving 
branch points
with the same coordinates but lying on different sheets is also still
valid and Fig.8b reproduces this fact correctly.

\vskip 15pt

\begin{tabular}{cc}
\psfig{figure=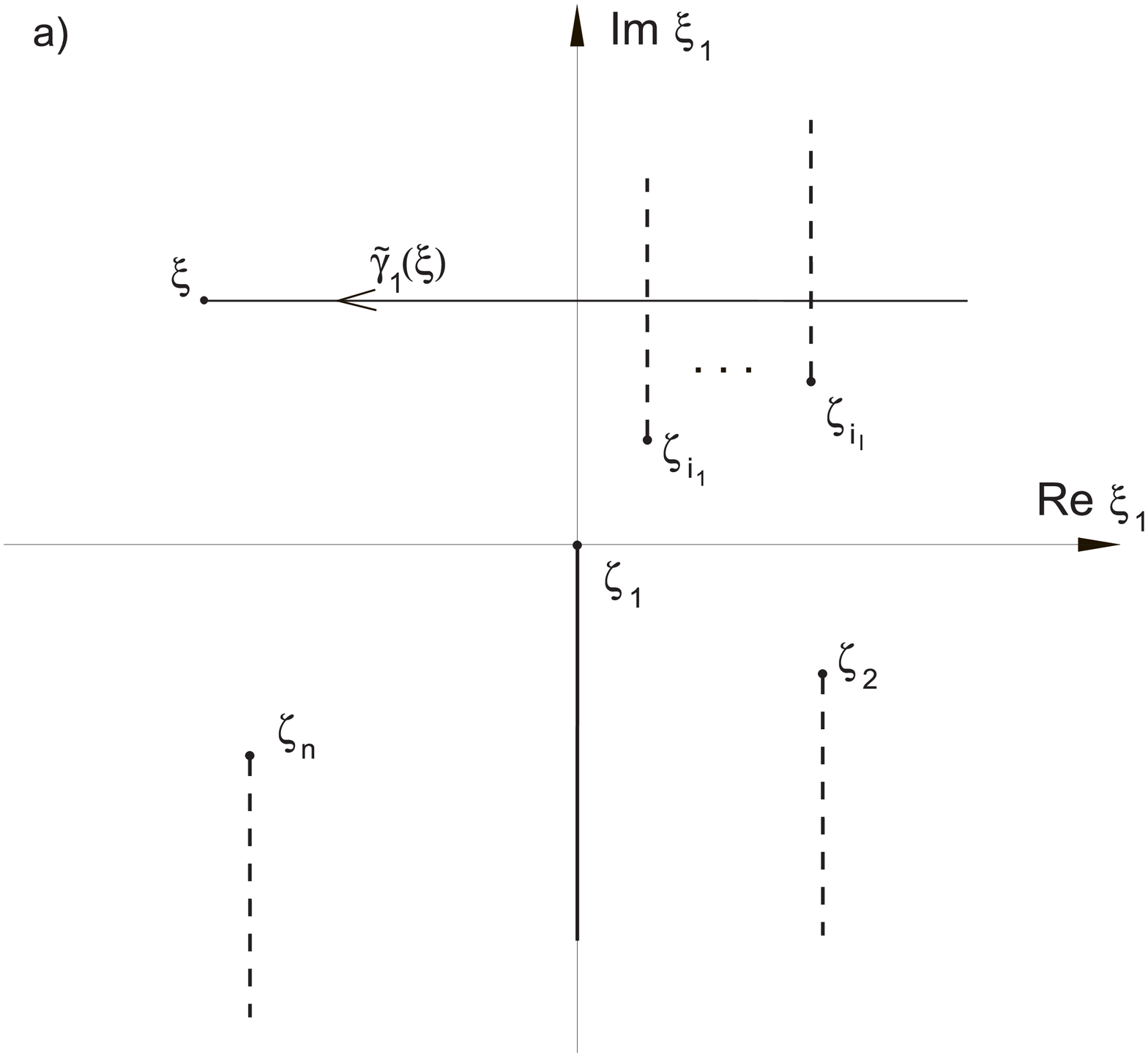,width=6.5cm} & \psfig{figure=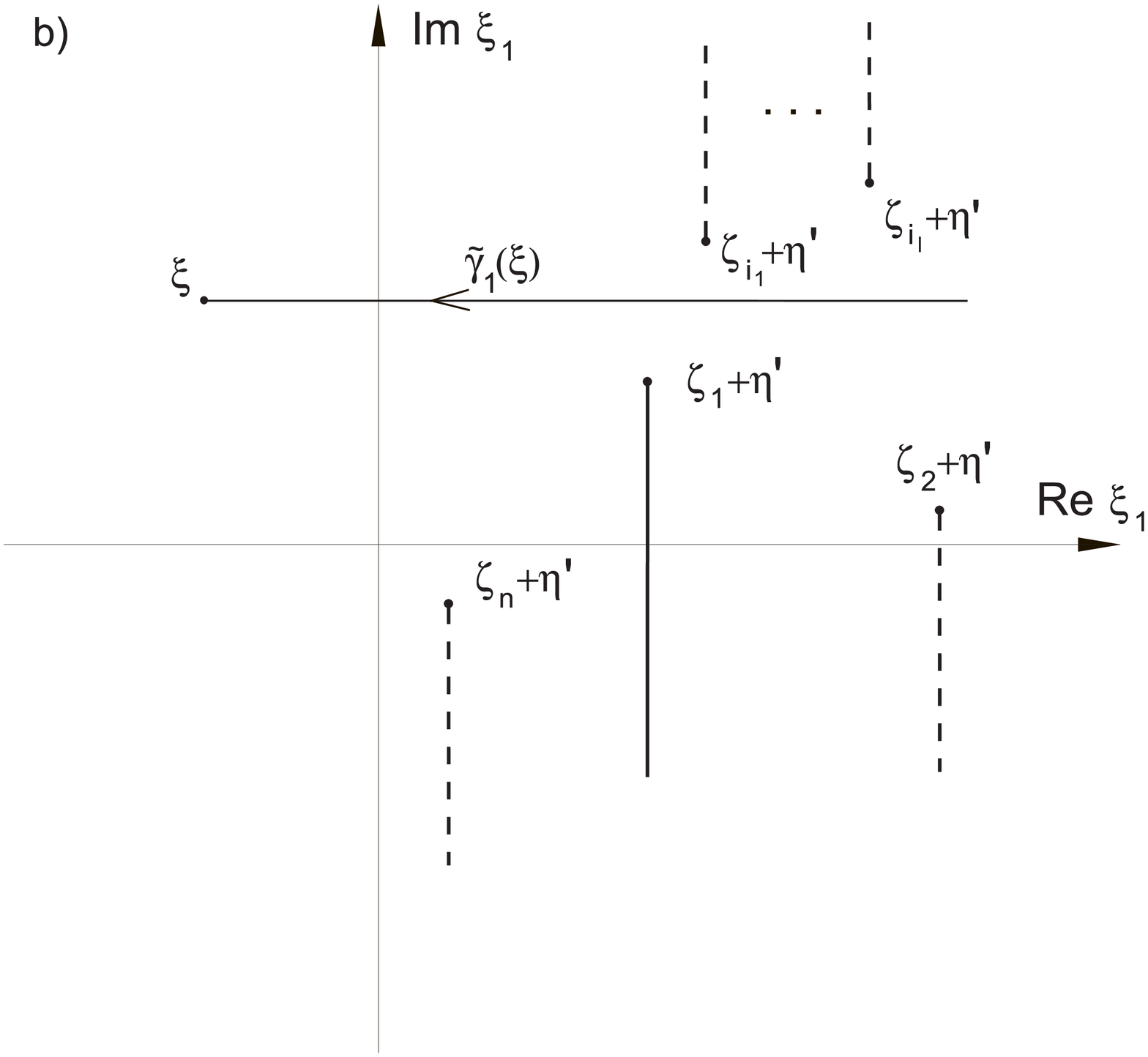,width=6.5cm} \\
Fig.9a The $\xi_1$-singularity structure of the&Fig.9b The
$\xi_1$-singularity structure of the\\only $\xi_1$-dependent factors of the integrand&integrand in the first formula (48)\\in the first formula (48) for $q=0$&for $q=0$
\end{tabular}

\vskip 15pt

\begin{tabular}{c}
\psfig{figure=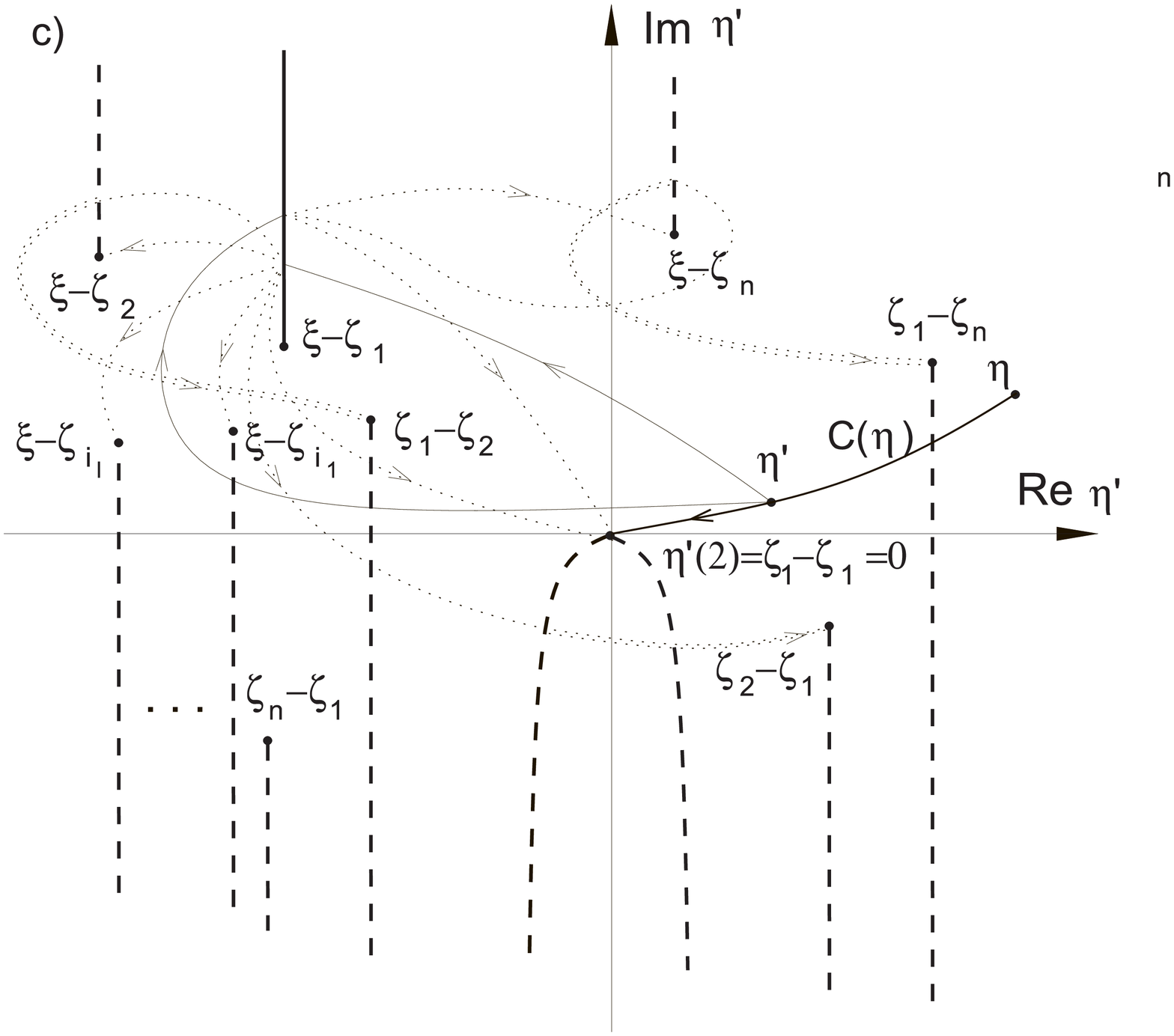,width=10cm} \\
Fig.9c The $\eta'$-singularity structure of the first formula (48)
  for $q=0$ \\after the $\xi_1$-integration
\end{tabular}

\vskip 15pt

Now consider the cases $\tilde{\Phi}^{(2q+1)}(\xi,s)$ and 
$\tilde{\Phi}^{(2q+2)}(\xi,s)$ assuming that the first sheets of the $\xi,s$
-Riemann surface structure of $\tilde{\Phi}^{(2q)}(\xi,s)$ is given by 
Fig.8b and Fig.9c (with the suitable $\eta,\eta'\to s$ substitutions
on the figures). It is clear that we can repeat
all the previous analyses and conclusions without any changes if the 
considered structure is
limited only to the sheets defined by the formulae (\ref{A1.14}) and the 
branch points shown on
the figures 8a, 9c, i.e. the structure on these figures is reproduced also 
for the 
functions $\tilde{\Phi}^{(2q+1)}(\xi,s)$ and $\tilde{\Phi}^{(2q+2)}(\xi,s)$. 
Since the series (\ref{A1.11}) is convergent this structure is the same also
 for $\tilde{\Phi}(\xi,s)$ itself.

It is important to note that the $fixed$ branch points on the $s$-Riemann 
surface are
generated in the scheme of the topological expansion (\ref{A1.11}) on $lower$ 
sheets.

Having $\tilde{\Phi}(\xi,s)$ defined in the above way we can restore 
the $\chi$-factors defined in the
sectors $S_1$ or $S_2$ by the Borel transformations of $\tilde{\Phi}(\xi,s)$ 
along the left halfaxis to get $\chi_1(x,\lambda)$ or
along the right one to get $\chi_2(x,\lambda)$. Clearly, the signs of 
$\lambda$ in both these integrations are different.

To prove the assertion of the theorem about the structure of the 
$\xi,s$-Riemann surface for $\tilde{\Phi}(\xi,s)$
continued to the sector $S_k$ along the contour $K'$ on Fig. 1a we should 
perform this continuation
on the $\xi,s$-Riemann surface corresponding to $\tilde{\Phi}(\xi,s)$ 
changing $\xi$ respectively and drawing cuts
properly. Such an operation, however, needs the detailed knowledge of the 
structure of many
lower sheets of the $\xi,s$-Riemann surface, a task which seems to be in 
general hopeless.
However, we have already noticed that such continuation can be easily 
performed with the help
of the formulae (\ref{A1.12})-(\ref{A1.14}) by changing the infinite ends of 
the $\xi$-integrations in these
formulae i.e. moving them to the appropriate sectors when the variable 
$\xi(=\xi(x))$ itself is
continued to these sectors when $x$ changes along the contour $K'$. For 
definitness let $\xi$ be
continued to the sector $S_3$. Then we can continue the mentioned infinite 
ends to the same sector.
According to the Stokes graph on Fig. 1a the corresponding pattern of the 
sheets on which $\tilde{\omega}(\xi)$ 
and $\Omega(\xi)$ are defined are shown on Fig.6. Therefore, 
taking this 
figure as the original one and
using again the formulae (\ref{A1.14}) we can repeat once again the analyses 
done above. The only
difference are introduced by additional branch points which can appear 
according to the Stokes
graph on Fig. 1a. Then forgetting about the branch points which lie on the 
lower sheets in Fig.6
(such as $\zeta_3$) we get as the final structure of the first sheets for 
$\tilde{\Phi}(\xi,s)$ the one shown in Fig.10.

The pattern on Fig.10 is the one which has also to follow if we would 
continue the
pattern of Fig.8b (with the $\eta\to s$ substitution) by moving anticlockwise the variable $\xi$ around the 
branch point $\zeta_1$ and
uprighting the cut emerging from this point by rotating it in the same 
anticlockwise direction.
Then the cut emerging from $\zeta_2$ on the figure is unscreened and 
uprighting it again anticloclwise
as well as all the other consecutive cuts met by $\xi$ on its way to the 
sector $S_3$ we have to reveal
by these {\it uprighting operations} of the cuts the pattern of Fig.10a. 
Compairing the latter with
Fig.8b we see that these operations do their job properly. The
proliferation of the branch points with the same coordinates on
different sheets plays an essential role in this operation allowing us
to recover the cuts which are screened by the uprighting operations
made over the cuts emerging from $\xi=\zeta_1$ and $\xi=\zeta_2$

On the $s$-Riemann surface of $\tilde{\Phi}(\xi,s)$ shown on Fig.9c
(with $\eta'\to s$ on the axes) the 
corresponding operation with
cuts are of course reversed i.e. each consecutive cut which is unscreened by 
the anticlockwise
upside-down rotation of its predecessor originating by the cut emerging from 
the branch point $\xi-\zeta_1$
 on Fig.9c has also to be rotated in the same way. 
This process stops on the cut emerging
from the last branch point at $\xi-\zeta_2$ being unscreened. We can then 
move this new pattern first
down leaving the origin of the first sheet to the right and next move to the 
right leaving the
origin of the sheet above all the moved branch points. The final position 
has to coincide then
with that shown in Fig.10b.

\vskip 15pt

\begin{tabular}{cc}
\psfig{figure=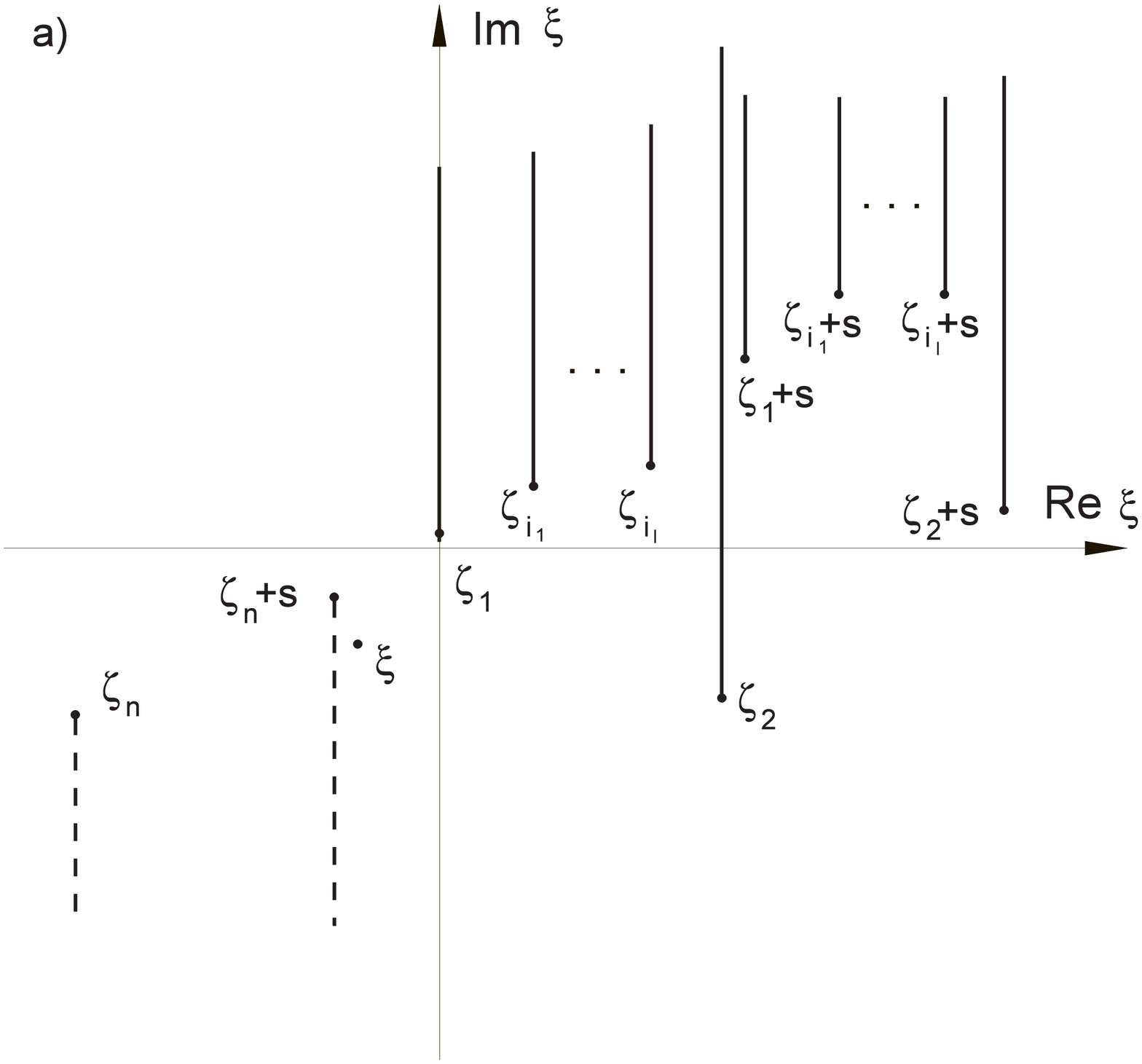,width=6.5cm} & \psfig{figure=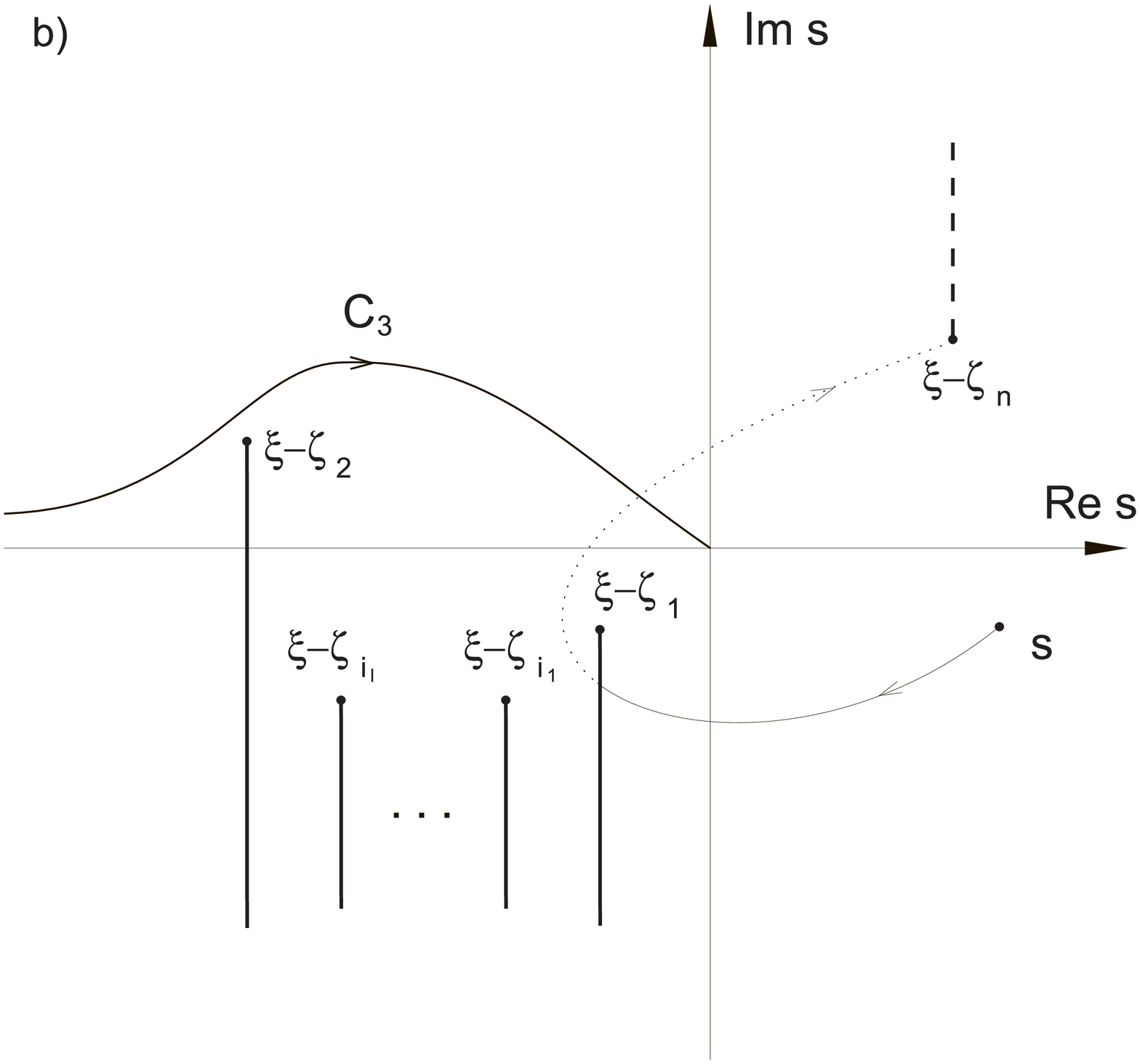,width=6.5cm} \\
Fig.10a The $\xi$-singularity structure 
   &Fig.10b The $s$-singularity structure \\
of $\tilde\Phi(\xi,s)$ for $x\in K'\cap S_3$&of $\tilde\Phi(\xi,s)$ for $x\in K'\cap S_3$
\end{tabular}

\vskip 15pt

We shall call the above operations with cuts leadind us to uncovering the 
desired sheets
of the $\xi,s$-Riemann surface the {\it unscreening operations}.

It is now clear that we can follow the above way considering the 
$\xi,s$-Riemann surface
structure corresponding to $\tilde{\Phi}(\xi,s)$ when $\xi$ is in $S_k$ 
being continued along the contour $K'$ on
Fig.1a. The tour of  along the contour $K'$ is mapped properly on the 
$\xi$-Riemann surface of $\tilde{\Phi}(\xi,s)$
 where $\xi$ moves anticlockwise avoiding all met branch points from the 
left and putting
upside-down the crossed cuts emerging from them realizing in this way the 
unscreening
operation. The same unscreening operations of putting the properly chosen 
cuts upside-down are
applied on the $s$-Riemann surface. We start from the pattern of 
Fig.9b and repeat the procedure
described above $k$ times. The final pattern has to have the form shown on 
Fig.11a,b. Its detailed
structure shown on the figure can be obtained from the formulae (\ref{A1.14})
 by the analyses described above.
\vskip 15pt
 
\begin{tabular}{cc}
\psfig{figure=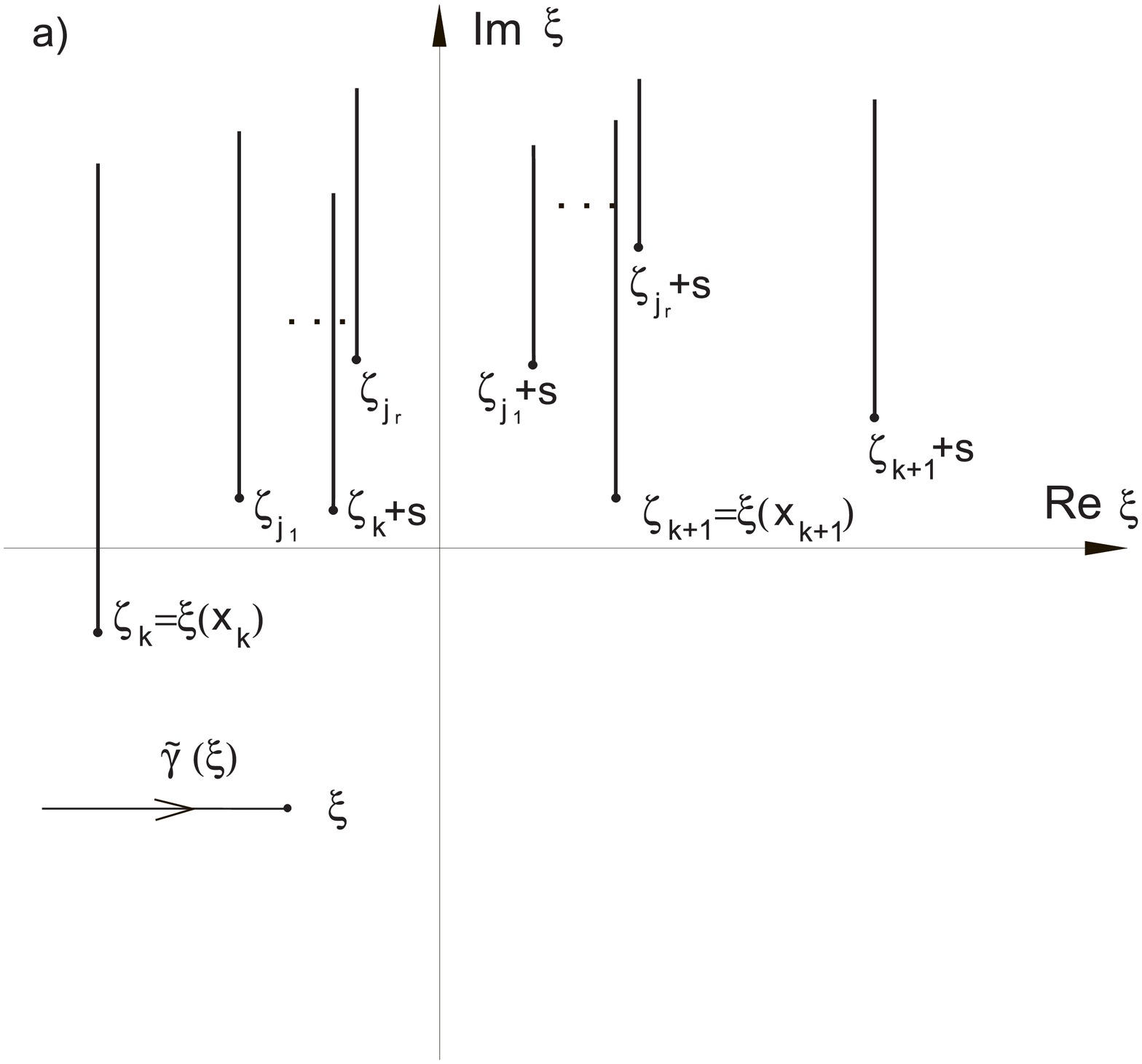,width=6.5cm} & \psfig{figure=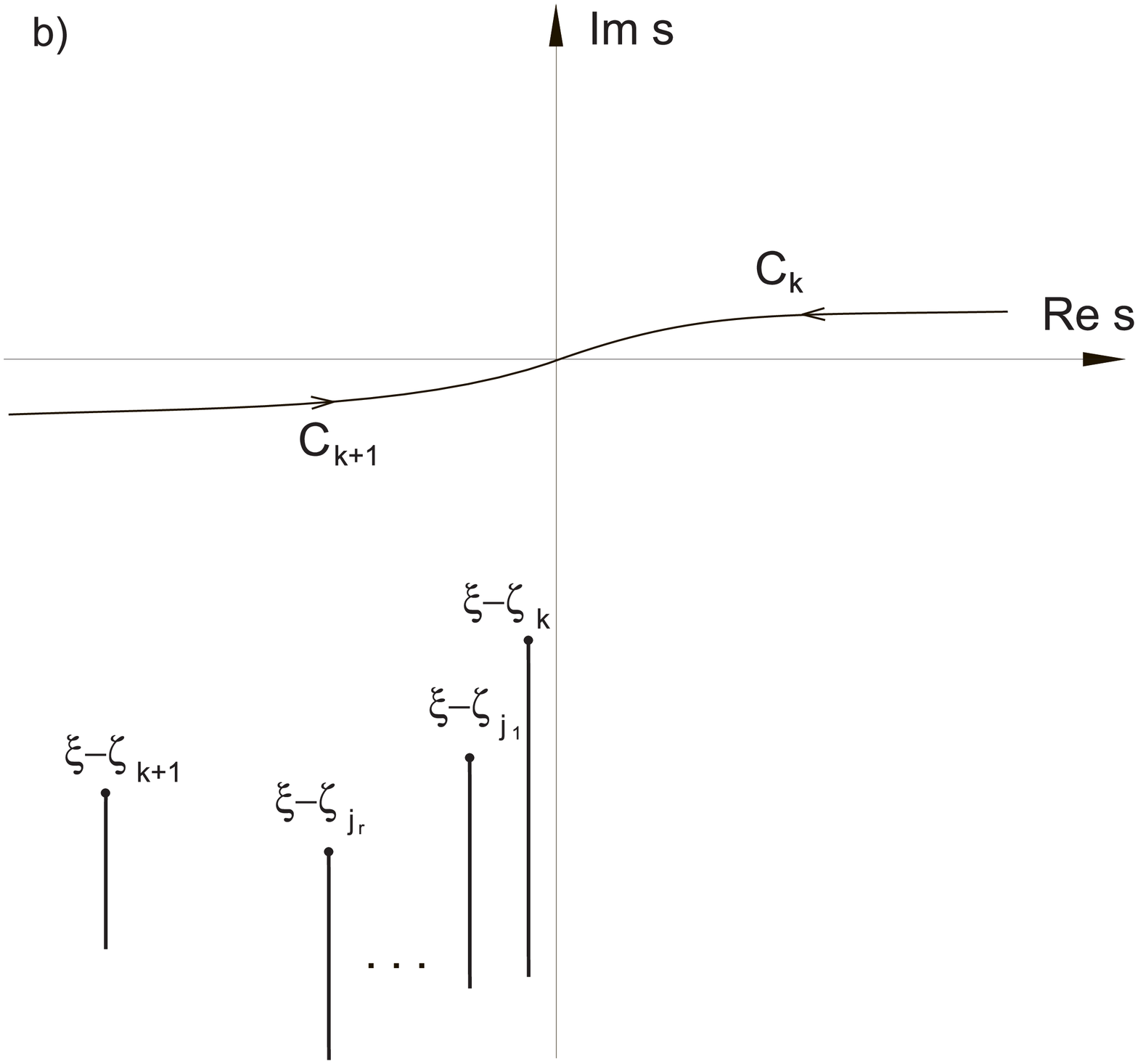,width=6.5cm} \\
Fig.11a The $\xi$-singularity structure &  Fig.11b The $s$-singularity structure\\of $\tilde\Phi(\xi,s)$
  for $x\in K'\cap S_k$&of $\tilde\Phi(\xi,s)$
  for $x\in K'\cap S_k$
\end{tabular}

\vskip 15pt

Borel transforming  along the left halfaxis of Fig.11b we recover the 
$\chi$-factor
corresponding to the sector $S_k$ (if $k$ is odd) or the one corresponding 
to the sector $S_{k+1}$ when the
Borel transformation is performed along the right halfaxis. QED.

To finish the above discussion let us note yet that as it follows from our 
estimation of
the convergence of the series (\ref{A1.11}) made in \cite{6} (see 
Appendix A.3 there) its $divergence$ on
each sheet of its $s$-Riemann surface (when $\xi$ is fixed) is no faster 
then exponential one.

\section*{Appendix 2}

\hskip+2em We shall show here that the Borel transformation (\ref{10})
of $\tilde{\chi}(x,s)$ (as given by (\ref{9}) with ${\chi}_{k,n}(x)$ in the 
latter satisfying the
reccurent relations (\ref{17}) (in its differential form)) along any
standard path satisfies the differential equation (\ref{15}).

To this end write (\ref{17}) in its differential form: 
\begin{eqnarray}
\chi_{n+1}^{\prime}(x) = q^{-\frac{1}{4}}(x) 
\left( q^{-\frac{1}{4}}(x) \chi_{n}(x) \right)^{\prime{\prime}} , \;\;\;\;\;\;
\;\;\;\;\; n \geq 0 \;\;\;\;\;\;\;\;
\label{42}
\end{eqnarray}

Next multiply both the
sides of (\ref{42}) by $(-s)^n /n!$ and sum them over $n$ ($n \geq 0$) to get:
\begin{eqnarray}
\frac{\partial^2 \tilde{\chi}(x,s)}{\partial s \partial x} + 
q^{-\frac{1}{4}}(x) \frac{\partial^2}{\partial x^2}
\left( q^{-\frac{1}{4}}(x) \tilde{\chi}(x,s) \right) = 0
\label{43}
\end{eqnarray}

Finally, multiply (\ref{43}) by $2\lambda e^{2\sigma\lambda s}$ and 
integrate (by parts) along a standard/cut path $C$ to have:
\begin{eqnarray}
2\sigma \lambda \left( 2 \lambda \int_C ds e^{2\sigma \lambda s}
\tilde{\chi}(x,s) \right)^{\prime} + 
q^{-\frac{1}{4}}(x) \left( q^{-\frac{1}{4}}(x) 
2 \lambda \int_C ds e^{2\sigma \lambda s}
\tilde{\chi}(x,s) \right)^{{\prime}\prime} = 0 \nonumber \\
\lambda > 0 \mbox{ , } \mbox{   } \mbox{   }
\sigma = 
\left\{
\begin{array}{lll}
+ 1 & \mbox{ for infinity of} & \Re C < 0 \\[5pt]
- 1 & \mbox{ for infinity of} & \Re C > 0
\end{array}
\right.
\label{44}
\end{eqnarray}

According to (\ref{10}) the equation (\ref{44}) coincides with (\ref{15}).

\vspace{6pt}

\section*{Appendix 3} 

\hskip+2em We shall show here that if $x$ is sufficiently close to
$x_0$ then the Borel function of the quotient of $\chi^{as}(x,\lambda)$ and
$\chi^{as}(x_0,\lambda)$ (with its factors corresponding to 
$\chi(x,\lambda)$ and $\chi(x_0,\lambda)$
respectively) can be integrated along the same standard path $\tilde{C}$
along which both the factors of the quotient can be summed too.
It means that all the three Borel functions, the quotient and
its two factors, are holomorphic in a common strip containing $\tilde{C}$.

To show this let us note that it is certainly true for the Borel
fuctions $\tilde{\chi}(x,s)$ and $\tilde{\chi}(x_0,s)$ 
of the two quotient factors considered separately
from the Borel function of the quotient itself.  This is the
result of the analytical dependence on $x$ of singularities of the
Borel functions of both these factors \cite{6}. Therefore there is a
strip $\tilde{S}$ on the Borel planes of $\tilde{\chi}(x,s)$
 and $\tilde{\chi}(x_0,s)$ containing a standard path $\tilde{C}$
along which these functions can be integrated to reproduce the
corresponding $\chi$-factors $\chi(x,\lambda)$ and $\chi(x_0,\lambda)$. 
It is now elementary to show that if $\tilde{\chi}(x_0,s)$ is 
holomorphic in $\tilde{S}$ then the Borel function of $\chi^{-1}(x_0,\lambda)$
 is also. This latter conclusion follows from the semiclassical 
expansion of $\chi^{-1}(x_0,\lambda)$. Namely, we have for this expansion
\be
\left( \frac{1}{\chi(x_0,\lambda)} \right)^{as} =
 \frac{1}{\chi^{as}(x_0,\lambda)}=
\sum\limits_{n \geq 0}  \frac{1}{C_0^{n+1}} \left(C_0 -
\chi^{as}(x_0,\lambda) \right)^n
\label{45}
\ee
where for $\chi^{as}(x_0,\lambda)$ we have assumed 
\be
\chi^{as}(x_0,\lambda) =
\sum\limits_{n \geq 0}  \frac{C_n}{(2\lambda)^n}
\label{46}
\ee
The expansion (\ref{45})
follows of course from the identical (in form) expansion of
$\chi^{-1}(x_0,\lambda)$ itself valid for $|\arg | \leq \pi /2$ 
when $\lambda$ is sufficiently large.  

The Borel function corresponding to the expansion (\ref{45}) is therefore
\be
\widetilde{ \frac{1}{\chi(x,\lambda)}} =
C_0 + \widetilde{ \sum\limits_{n \geq 1}  
\frac{(C_0 - \chi(x,\lambda) )^n}{C_0^{n+1}} } =
C_0 + \sum\limits_{n \geq 1}  
\frac{(C_0 - \tilde{\chi})^{*n} (x,s)}{C_0^{n+1}}  \nn 
\\
\label{47}
= C_0 + \sum\limits_{n \geq 1} 
\frac{(-1)^{n+1}}{C_0^{n+1}}
\int\limits_0^s ds_1 (C_0 - \tilde{\chi}(x,s-s_1))
\int\limits_0^{s_1} ds_2 \tilde{\chi}'(x,s_1 -s_2) ... 
\\
... \int\limits_0^{s_{n-3}} ds_{n-2} \tilde{\chi}'(x,s_{n-3} -s_{n-2})
\int\limits_0^{s_{n-2}} ds_{n-1} \tilde{\chi}'(x,s_{n-2} -s_{n-1}) \nn
\ee
where the prime at $\tilde{\chi}'(x,s)$ means the differentiation over $s$ and
where the following definition of the star (convolution)
operation has been used 
\be
(\tilde{f} * \tilde{g}) (s) = \frac{d}{ds}
\int\limits_0^{s} \tilde{f}(s) \tilde{g}(s -s') ds'
\label{48}
\ee

From the representation (\ref{47}) it follows easily that the strip
$\tilde S$ of the holomorphicity of $\tilde{\chi}(x,s)$ is also such a
strip for the Borel function of $\chi^{-1}(x,\lambda)$ since the series in (\ref{47}) is uniformly convergent in $\tilde{S}$.

\end{document}